%% file: full-document.tex
\documentclass[conference]{IEEEtran}
\usepackage{epsfig,endnotes}

\usepackage{epsfig}     
\usepackage{epstopdf}

\usepackage{graphicx}
\usepackage{caption}
\usepackage{subcaption}
\usepackage{xspace}

\usepackage{multirow}
\usepackage{array}
\usepackage{color}
\usepackage{times}
\usepackage{tablefootnote}
\usepackage{tabularx}
\usepackage{balance}
\usepackage{url}

\newcommand{\zebra}{ZEBRA\xspace}
\newcommand{\KBactivity}{keyboard-only\xspace}
\newcommand{\allactivity}{all-activity\xspace}
\newcommand{\attacker}{$\mathcal{A}$\xspace}
\newcommand{\victim}{$\mathcal{V}$\xspace}
\newcommand{\attackedterminal}{$\mathcal{AT}$\xspace}
\newcommand{\victimdevice}{$\mathcal{VD}$\xspace}

\newfont{\eaddfnt}{phvr8t at 10pt}

\newif\ifabridged
\newif\ifanonymous
\newif\ifcomments
\newif\ifllncs
\newif\ifapps

\ifdefined\isapps
\appstrue
\fi

\ifdefined\isllncs
\llncstrue
\fi

\ifdefined\isabridged
\abridgedtrue
\fi

\ifdefined\isanonymous
\anonymoustrue
\fi

\ifdefined\iscomments
\commentstrue
\fi

\ifcomments
\newcommand\otto[1]{\textcolor{red}{Otto: #1}}
\newcommand\asokan[1]{\textcolor{green}{Asokan: #1}}
\newcommand\mika[1]{\textcolor{blue}{Mika: #1}}
\newcommand\changeOtto[1]{\textcolor{red}{#1}}
\newcommand\changeMika[1]{\textcolor{red}{#1}}
\newcommand\changeAsokan[1]{\textcolor{red}{#1}}
\else

\newcommand\otto[1]{}
\newcommand\asokan[1]{}
\newcommand\mika[1]{}
\newcommand\added[1]{{#1}}

\newcommand\changeOtto[1]{{#1}}
\newcommand\changeMika[1]{{#1}}
\newcommand\changeAsokan[1]{{#1}}
\fi

\newcommand{\figwidth}{\columnwidth}

\begin{document}

\title{Pitfalls in Designing Zero-Effort Deauthentication: Opportunistic Human Observation Attacks}

\ifanonymous
\else
\author{\IEEEauthorblockN{Otto Huhta\IEEEauthorrefmark{1}, Prakash Shrestha\IEEEauthorrefmark{2}, Swapnil Udar\IEEEauthorrefmark{1}, Mika Juuti\IEEEauthorrefmark{1},
Nitesh Saxena\IEEEauthorrefmark{2} and N. Asokan\IEEEauthorrefmark{3}}
\IEEEauthorblockA{\IEEEauthorrefmark{1}Aalto University}
\IEEEauthorblockA{\IEEEauthorrefmark{2}University of Alabama at Birmingham}
\IEEEauthorblockA{\IEEEauthorrefmark{3}Aalto University and University of Helsinki}
\IEEEauthorblockA{\{otto.huhta, swapnil.udar, mika.juuti\}@aalto.fi, \{saxena, prakashs\}@uab.edu, asokan@acm.org}}

\fi

\IEEEoverridecommandlockouts
\makeatletter\def\@IEEEpubidpullup{9\baselineskip}\makeatother
\IEEEpubid{\parbox{\columnwidth}{Permission to freely reproduce all or part
    of this paper for noncommercial purposes is granted provided that
    copies bear this notice and the full citation on the first
    page. Reproduction for commercial purposes is strictly prohibited
    without the prior written consent of the Internet Society, the
    first-named author (for reproduction of an entire paper only), and
    the author's employer if the paper was prepared within the scope
    of employment.  \\
    NDSS '16, 21-24 February 2016, San Diego, CA, USA\\
    Copyright 2016 Internet Society, ISBN $<$TBD$>$\\
    http://dx.doi.org/10.14722/ndss.2016.23199
}
\hspace{\columnsep}\makebox[\columnwidth]{}}

\maketitle

\input{chapters/abstract}

\input{chapters/introduction}

\input{chapters/2background}

\input{chapters/3-1revisiting_zebra}

\input{chapters/3-2attack_overview}

\input{chapters/4system_design}


\input{chapters/6results}

\input{chapters/discussion}

\input{chapters/related_work}

\input{chapters/conclusions}

\ifanonymous
\else
\vspace{3mm}
\noindent\textbf{Acknowledgments}: This work was supported in part by
the Academy of Finland ``Contextual Security'' project (274951),
NSF grant CNS-1201927 and a Google Faculty Research Award. 
We thank Shrirang Mare for explaining \zebra design parameters, and
Hien Truong and Babins Shrestha for discussions on
transparent deauthentication.
\fi

\balance

{\footnotesize \bibliographystyle{plain}
\bibliography{etc/references}}

\input{chapters/appendices}


\end{document}

%% file: chapters/abstract.tex
\begin{abstract}

	Deauthentication is an important component of any authentication
	system. The widespread use of computing devices in daily life has
	underscored the need for \textit{zero-effort} deauthentication schemes.
	However, the quest for eliminating user effort may lead to hidden
	security flaws \ifabridged \else in the authentication schemes\fi. 

As a case in point, we investigate a prominent zero-effort deauthentication
scheme, called \zebra, which provides an interesting and a useful solution to a
difficult problem as demonstrated in the original paper. We identify a subtle
incorrect assumption in its adversary model that leads to a fundamental design flaw. We
exploit this to break the scheme with a class of attacks that are much easier
for a human to perform in a realistic adversary
model, compared to the na\"{i}ve attacks studied in the \zebra paper. For
example, one of our main attacks, where the human attacker has to opportunistically
mimic only the victim's keyboard typing activity at a nearby terminal, is
significantly more successful compared to the na\"{i}ve attack
that requires mimicking keyboard and mouse activities as well as keyboard-mouse movements.
  Further, by understanding the design flaws in \zebra as cases of \textit{tainted
  input}, we show that we can draw on well-understood design principles to
  improve \zebra's security. 
  
  \end{abstract}

%% file: chapters/introduction.tex
\section{Introduction}
\label{sec:intro}

User authentication is critical to many on-line and off-line services. 
\ifabridged
\else
Computing devices of all types and sizes,
ranging from mobile phones through personal computers to remote
servers rely on user authentication.  
\fi
\textit{Deauthentication} --
promptly recognizing when to terminate a previously authenticated user
session -- is an essential component of an authentication system.
\ifabridged
\else

\fi
The pervasive use of computing in people's daily lives underscores the
need to design effective, yet intuitive and easy-to-use deauthentication
mechanisms. However, this remains an important unsolved problem in
information security. A promising approach to improving usability
of (de)authentication mechanisms is to make them \textit{transparent} to
users by reducing, if not eliminating, the cognitive effort required
from users.  Although such \textit{zero-effort} authentication schemes
are compelling, designing them correctly is difficult. The need
to minimize additional user interactions required by the
scheme is a severe constraint that can lead to design
decisions which might affect the security of the scheme.

One prominent approach for improving usability of security mechanisms
involves comparing information observed from two different sources. Such
a \textit{bilateral} approach has been proposed as part of solutions
for a variety of security problems such as deauthentication of
users~\cite{mare2014zebra}, determining if two or more devices are
co-present in the same place~\cite{TruongPerCom14}, establishing
security associations among nearby devices
(``pairing'')~\cite{VanDyken+14,DBLP:journals/csur/ChongMG14} and
authorizing transactions between co-present devices~\cite{bump}.
Bilateral authentication schemes are attractive because they can avoid
imposing any cognitive load on users (thus making them
``zero-effort''), or the need to store sensitive or user-specific information on 
devices~\cite{mare2014zebra}.  However, an adversary \changeAsokan{capable of influencing} one or
both sources of information being compared in a bilateral 
scheme may compromise security.

In this paper, we illustrate the problem of subtle flaws in the design
of zero-effort bilateral schemes by examining an interesting class of schemes \changeAsokan{represented by \zebra,} a zero-effort bilateral deauthentication scheme, proposed
recently in a premier security research venue~\cite{mare2014zebra}.
\zebra is intended for scenarios where users authenticate to ``terminals'' (such as \changeAsokan{desktop computers}). In such scenarios, users typically have
to either manually deauthenticate themselves by logging out or locking the
terminal, or the terminal can deauthenticate a user automatically after
a sufficiently long period of inactivity. The former requires user
effort while the latter sacrifices promptness. \zebra attempts to make
the process of deauthentication \textit{both prompt and transparent}: once a
user is authenticated to a terminal (using say a password), \changeAsokan{it} continuously,
yet transparently \textit{re-authenticates} the user so that prompt
deauthentication is possible without explicit user action. A user is required to wear a bracelet equipped with sensors on his
\changeMika{mouse-holding} hand. The bracelet is wirelessly connected to
the terminal, which compares the sequence of events it observes
(e.g., keyboard/mouse interactions) with the sequence of events
inferred using measurements from the bracelet
sensors. \changeAsokan{The logged-in user is deauthenticated} when the two sequences no
longer match. 

\zebra is particularly compelling because of its
simplicity of design. 
\changeAsokan{However, the simplicity hides a design assumption that an adversary can exploit to defeat the scheme. We show how a more realistic adversary can circumvent ZEBRA. Since no implementation of ZEBRA was available, we built an end-to-end implementation and use it in our attack. We also implemented changes needed to make ZEBRA work in real-time.}

\begin{figure*}[!bht]
\centering
\includegraphics[width=.8\textwidth]{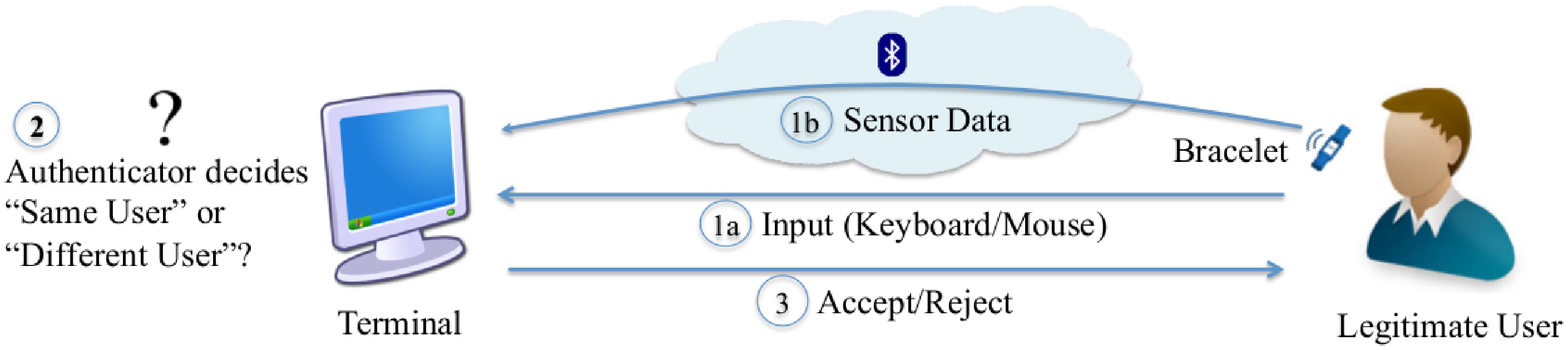}
\caption{Normal operation of \zebra}
\label{fig:benign}
\end{figure*}

Our primary contributions can be summarized as follows:
\begin{enumerate}
\itemsep0em
\item 
We highlight fundamental pitfalls in designing zero-effort bilateral
security schemes by studying \zebra, a notable prior scheme. We 
identify a hidden design choice in \zebra that allows us to develop an \textbf{effective attack strategy}: a human attacker observing a victim at a nearby terminal and \textit{opportunistically} mimicking only a subset of the victim's activities (e.g., keyboard events) at the authentication terminal (Section~\ref{sec:attack_overview}).
\itemsep0.5em
\item  We build a \textbf{end-to-end
    implementation}\footnote{Unlike \cite{mare2014zebra} which only
    described the implementation of individual components \changeAsokan{and off-line classification}.} of \zebra
  (Section~\ref{sec:system_setup}), and demonstrate via \textbf{experiments in
  realistic adversarial settings} that \zebra as designed can be
  defeated by our opportunistic attacker with a \changeMika{(statistically)} significantly higher probability compared to a \changeAsokan{na\"{i}ve} attacker, \changeMika{also} considered in \cite{mare2014zebra} (one who attempts to mimic all, keyboard and mouse, activities) (Section~\ref{sec:results}).
\itemsep0.5em
\item We cast \zebra's design flaw as a case of
  \textbf{tainted input}, and thus draw from well-understood principles
  of secure system design that may help improve the security of \zebra (Section~\ref{subsec:strengthening}).
\end{enumerate}

%% file: chapters/2background.tex
\section{Background}
\label{sec:background}

\ifabridged
\else
Since we use \zebra~\cite{mare2014zebra} as our exemplary bilateral
zero-effort deauthentication scheme, we now describe it in more
detail. 
\fi
\changeAsokan{It} is intended for multi-terminal environments where users frequently move between terminals. Mare et al.~\cite{mare2014zebra} present a hospital environment as their motivating scenario. Hospital staff members often use shared terminals. However, a user must not, intentionally or unintentionally, access hospital systems from terminals where other users have logged in. 
Users may leave terminals without logging out, but may still remain in the vicinity. Proximity-based zero-effort deauthentication schemes such as ZIA~\cite{CN02} or BlueProximity~\cite{Blueproximity} cannot be used because these methods are not accurate enough for short distances. Although the motivating scenario is an environment with shared terminals, zero-effort deauthentication schemes \ifabridged
\else
like \zebra
\fi
are broadly applicable to any scenario where users may leave their terminals unattended.

\vspace{2mm}
\noindent\textbf{Adversary Model}:
\zebra \changeMika{\cite{mare2014zebra}} considers two types of adversaries: ``innocent'' and ``malicious''. An innocent adversary is a legitimate user who starts using an unattended terminal inadvertently without realizing that another user (``victim'') is logged into that terminal. In contrast, a malicious adversary deliberately uses an unattended terminal of the victim with the intent of performing some action impersonating the victim. A malicious adversary may observe the behavior and actions of the victim (such as imitating the victim's hand movements \changeAsokan{made} while interacting with another terminal). The goal of \zebra is to quickly detect if a previously authenticated session on a terminal is being used by anyone other than the user who originally authenticated, and promptly deauthenticate the session. Naturally, decisions made by \zebra should minimize false positives
(incorrectly recognizing an adversary as the original authenticated user, thereby failing to deauthenticate him as well as false negatives (incorrectly concluding that current user is not the original user, thereby deauthenticating him.

\vspace{2mm}
\noindent\textbf{System Architecture}:
Figure~\ref{fig:benign} depicts the normal (benign) operation of \zebra.
It correlates a user's activities on a terminal with measurements of user activity relayed from a wrist-worn device (we call it a bracelet for simplicity, but it can be a general-purpose smartwatch as in our implementation and analysis). 
\ifabridged
\else
The goal is to continuously verify that the logged in user is the one using the terminal and to quickly deauthenticate any unintended users. 
\fi
\zebra assumes terminals with keyboard/mouse and a personal bracelet for each user of the system. The bracelet has accelerometer and gyroscope sensors to record wrist movements. Terminals and bracelets securely communicate using ``paired'' wireless channels like Bluetooth. In addition, a terminal knows the identity of the bracelet associated with each authorized user. Users initially authenticate themselves to terminals using some mechanism external to \zebra (such as using a username/password). Once a user has been authenticated, the terminal connects to \changeAsokan{that} user's bracelet and starts receiving sensor measurements from it.

\begin{figure*}[!htb]
\centering
\vspace{-8mm}
\includegraphics[width=0.9\textwidth]{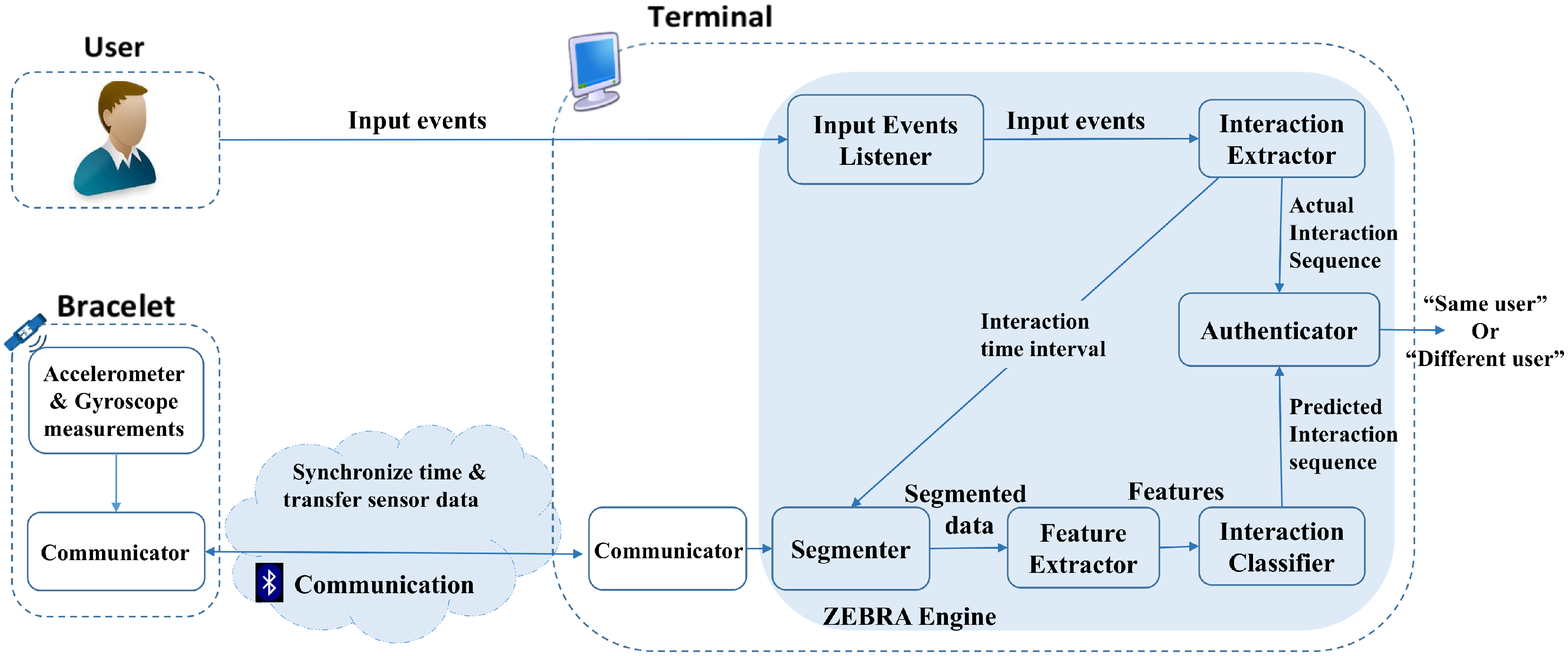}
\vspace{-20mm}
\caption{\zebra system architecture}
\label{fig:architecture}
\vspace{-5mm}
\end{figure*}

The basic principle of operation is to compare the sequence of user activity seen at the terminal with that inferred from data sent by the bracelet. \zebra's system architecture is shown in Figure \ref{fig:architecture}. An \textit{Interaction Extractor} on the terminal identifies the \textit{actual interaction sequence} based on input events observed by the terminal peripherals. \changeAsokan{It} defines three different types of such interactions: typing, scrolling, and hand movements between the mouse and keyboard (referred to as ``MKKM'')\footnote{\added{\zebra neither cares about which key was pressed nor about what direction the mouse was scrolled. It actually cares about whether the interaction is typing, scrolling or movement between mouse and keyboard.}}.  Interaction Extractor records the timestamps of each event in the actual interaction sequence.
A \textit{Segmenter} on the terminal receives measurement data sent by the bracelet
and segments this data according to the timestamps it receives from Interaction Extractor.
Segmenter ignores all measurements that fall outside these time slots. 
From the segments, a \textit{Feature Extractor} extracts salient features and feeds them to an \textit{Interaction Classifier} that has been trained to identify the type of interaction from bracelet measurement data.
The classifier outputs a \textit{predicted interaction sequence}. Finally, an \textit{Authenticator} compares the two interaction sequences and determines whether the current user at the terminal is the ``same'' as, or ``different'' from, the originally authenticated user.

Authenticator can be tuned by a number of parameters. It compares sequences of length \textit{w (window size)} at a time. In each window, if the fraction of matching interactions exceeds a threshold \textit{m (matching threshold)}, it records 1 for that window; otherwise it records 0. If the record is 0 for \textit{g (grace period)} successive windows, the authenticator outputs ``different'' causing \zebra to deauthenticate the session. Successive windows may overlap, as determined by \textit{f (overlap fraction)}, with 0 signifying no overlap.

Segmenter ignores readings from the bracelet when Interaction Extractor detects no activity on the terminal. \changeMika{The choice was motivated in} \changeAsokan{\cite{mare2014zebra} by privacy considerations: the user's activities are} \changeMika{not monitored when nobody is using the terminal.} \changeAsokan{At first glance, it} is a natural and reasonable design decision: if there is no terminal activity, there is reason to deauthenticate the session (thus reducing the chances of false negative decisions). However, as we shall see, an adversary can exploit this subtle aspect of the design.

\vspace{1mm}
\noindent\changeMika{\textbf{Validation}: Mare et al.~\cite{mare2014zebra} validated usability of their deauthentication scheme by calculating false negative rates for normal usage scenarios with different parameter settings. They validated the security by considering three separate scenarios.} \changeAsokan{The first two scenarios model the ``innocent adversary'':} \changeMika{the logged in user (victim) is either walking or writing nearby while the} \changeAsokan{attacker accesses the victim's terminal. The last scenario models the ``malicious'' adversary: the victim uses} \changeMika{\emph{another} terminal, while the} \changeAsokan{attacker uses the victim's original terminal. The activity conducted by both victims and attackers is filling forms. These scenarios were chosen as representative of} \changeMika{multi-user environments such as hospitals, where physicians enter form-type data about their patients and routinely forget to log out of their terminals. 
It is reasonable to assume there are multiple terminals that} \changeAsokan{users access and use.}
\changeMika{Similar usage scenarios are plausible in other contexts as well, such as in factory floors or control rooms. In \cite{mare2014zebra}, the} \changeAsokan{malicious adversary is required} \changeMika{to mimic \emph{all} mouse-hand movements of the victim. Ordinary non-expert users act as the attackers in their analysis.} \changeAsokan{Because Mare et al. \cite{mare2014zebra}} \changeMika{``realize that a real adversary can be motivated and skilled enough to mimic user very well, compared to our adversaries'',} \changeAsokan{they tried to make the scenario advantageous to the attacker by (a) providing the attacker with a clear view of the victim's screen and (b) have the victim give verbal cues to indicate what the victim was doing during the experiments (e.g., answering question 2 in the form).} 
\changeMika{They concluded that their system was able to deauthenticate such attackers in reasonable time, while keeping false negative rates low. 
}



%% file: chapters/3-1revisiting_zebra.tex
\section{Our Attack}
\label{sec:attack_overview}


There are a number of attributes that make \zebra attractive. In
particular, rather than trying to \textit{recognize} the user,
\zebra's bilateral approach simply \textit{compares} two sequences
that characterize user interaction. Consequently, its decisions neither limit how a user interacts with the terminal nor require storing any information about the user or his style of interaction. Such simplicity makes \zebra robust but also vulnerable. In this section, we revisit the security analysis in \cite{mare2014zebra}, point out a design flaw, and explain how it can be used to attack \zebra.

\subsection{Revisiting \zebra Security Analysis}
\label{sec:revisiting_zebra}

Recall from Section~\ref{sec:background} that Segmenter ignores all measurement data from the bracelet during periods when Interaction Extractor does not record any activity \changeAsokan{on the terminal involving the three types of interactions recognized by \zebra}. 
However, the attacked terminal is under the control of the adversary \changeMika{and thus she} can effectively choose which parts of the bracelet measurement data will be used by \zebra\changeMika{ to re-authenticate the user}. 
Mimicking all interactions is not the best \changeAsokan{attack strategy}. 
A smart adversary can selectively choose only a subset of the victim's interactions to mimic since it can ensure that the rest of the victim's interactions will be ignored by \changeAsokan{Authenticator}. \changeOtto{Furthermore,} \changeMika{to validate security,} 
\changeAsokan{we need to use a realistic adversary model which allows attackers to be skilled and experienced in mimicking how people interact with terminals. It is unreasonable to use inexperienced test participants to model the adversary.} 
Thus, the role of the attacker in this paper was played by two members of our research group that were knowledgeable of the \zebra system and experienced at mimicking attacks.

%
%


%% file: chapters/3-2attack_overview.tex
\subsection{Attack Scenarios and Strategies}
\label{subsec:scenarios}



\begin{figure*}[!htbp]
\centering
\vspace{-5mm}
\includegraphics[width=.95\textwidth]{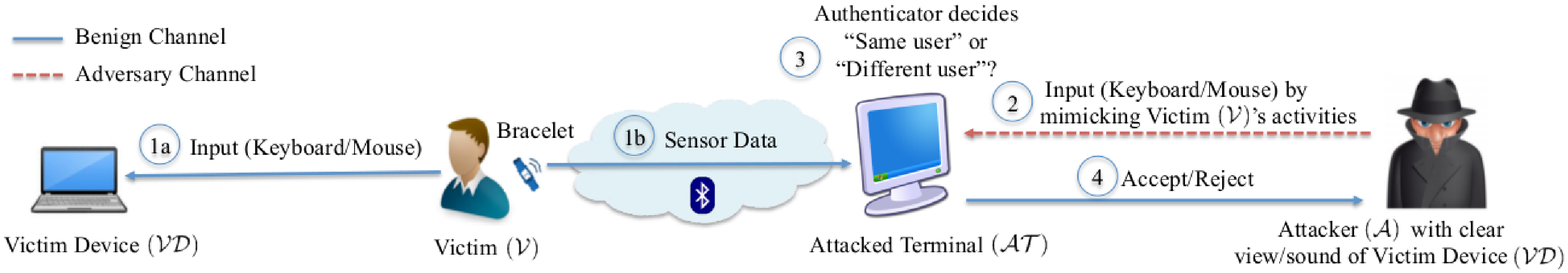}
\vspace{-3mm}
\caption{Basic attack setting}
\label{fig:rta-scenario}
\end{figure*}

In our attack scenarios, we model a malicious adversary against \zebra
as discussed in Section~\ref{sec:background}.  We assume that the
adversary \attacker accesses the attacked terminal \attackedterminal when the victim \victim steps away from
it without logging out. We also assume that \victim is using
another computing device (the ``victim device'', \victimdevice) elsewhere (e.g., a nearby terminal). \changeAsokan{Figure~\ref{fig:rta-scenario} illustrates the attack setting.}

\vspace{1mm}
\noindent\textbf{Strategy}: The goal of \attacker is to remain logged in on \attackedterminal for as long as possible, while interacting with the terminal. To this end, \attacker needs to consistently produce a sufficiently large fraction of interactions that will match \victim's interactions on \victimdevice. Since \attackedterminal is under the control of \attacker, it can choose when \attackedterminal's Interaction Extractor triggers Authenticator to compare the predicted and actual interaction sequences. If \attacker adopts an \textit{opportunistic} strategy, it can \textit{selectively} choose only a subset of \victim's interactions to mimic so as to maximize the fraction of matching interactions.
We conjecture that such an opportunistic adversary will be more
successful than the na\"{i}ve adversary that was considered
in~\cite{mare2014zebra}.

First, we consider a
\textit{\KBactivity} attack where \attacker mimics only the typing
interactions while ignoring all others. Typing sequences are typically longer and less prone to delays in mimicking.
The opportunistic strategy is for \attacker to start typing only after \victim starts
typing and attempt to stop as soon as 
\victim stops. A sophisticated \KBactivity attacker may estimate
the expected length of \victim's typing session and
attempt to stop before \victim does. If \attacker makes just a few key presses each time \victim begins typing, he can be confident that the actual interaction sequence he produces will match the predicted interaction sequence. These \KBactivity attacks are powerful because in all modern personal computer operating systems a wide range of actions can be performed using only the keyboard.

Second, we consider an \textit{\allactivity} attack, where 
\attacker mimics
all types of interactions (typing, scrolling and MKKM) but
opportunistically chooses a subset of the set of interactions.
\ifllncs
This opportunistic strategy is described in detail in Appendix~\ref{app:extra_attackers}.
\else
As before, the \attacker's selection
criterion is the likelihood of correctly mimicking \victim. In
particular, \attacker will use the following strategy:
\begin{itemize}
\itemsep0em
\item Once \attacker successfully mimics a keyboard to mouse interaction, he
  is free to carry out any interaction involving the mouse (scroll,
  drag, move) at will because the bracelet measurements for all
  \changeAsokan{interactions} involving the mouse are likely to be similar.
\item If \attacker fails to quickly mimic a keyboard to mouse (or vice
  versa) interaction, he
  does nothing until the next opportunity for an MKKM
  interaction arises (foregoing all interactions until after the MKKM
  is completed).
\end{itemize}
\fi

\zebra concatenates continual typing events into up-to 1 second long interactions: as such the typing speed of \attacker is not particularly relevant. Instead, \attacker may divert more of his attention to observing \victim.

\vspace{1mm}
\noindent\textbf{Observation Channels}: By default, and similar to \cite{mare2014zebra}, we consider an adversary \attacker who has a
clear view of \victim's \changeAsokan{interactions} (Figure~\ref{fig:rta-scenario}). This models two cases: where \attacker has direct visual access to \victim and where \attacker has access to a \textit{video aid} such as a surveillance camera aimed at \victimdevice. During our attacks that use visual information of the victim's behavior, victim's new device \victimdevice was placed next to the victim terminal \attackedterminal. We also consider the case where \attacker has no visual access to but can still hear sounds resulting from \victim's activities. Again, this models two cases: where both \victim and \attacker are in the same physical space separated by a visual barrier (e.g., adjacent cubicles) and where \attacker has planted an \textit{audio aid} (e.g., a small hidden bug or a microphone) close to \victimdevice. 

\vspace{1mm}
\noindent\textbf{Scenarios}: The combination of attack strategy and
type of observation channel leads to several different attack scenarios. We consider four of the most significant ones:

\begin{itemize}
\itemsep0em
\item In \textbf{\changeMika{na\"ive} \allactivity} attack, \attacker is able to both see and hear \victim. \attacker attempts to mimic \textit{all interactions} of \victim. This is the attack scenario proposed and studied in \cite{mare2014zebra}.
\itemsep0.35em
\item In \textbf{opportunistic \KBactivity} attack, \attacker is able to both see and hear \victim. \attacker selectively mimics only a \textit{subset of \victim's typing interactions}.
\itemsep0.35em
\item In \textbf{opportunistic \allactivity} attack, \attacker is able to both see and hear \victim. \attacker selectively mimics a \textit{subset of all types of interactions} of \victim following the guidelines mentioned 
\ifllncs
in Appendix~\ref{app:extra_attackers}.
\else
above.
\fi
\itemsep0.35em
\item In \textbf{audio-only opportunistic \KBactivity} attack, \attacker is able to hear, but not see, \victim's \changeAsokan{interactions}. \attacker listens for keyboard activity and attempts to mimic \textit{a subset of \victim's typing interactions}.

\end{itemize}
	
	While one can imagine other attack combinations, we consider these four to be representative of different choices available to \attacker. For example, we leave out an audio-only \allactivity attack because it is unlikely to succeed. Although our experiments are ``unaided'' (i.e., no audio or video recording), the results generalize to aided scenarios, if data transmission between the aid and the attacker does not introduce excessive delays.

%% file: chapters/4system_design.tex
\section{\zebra End-to-End System}
\label{sec:system_setup}

Mare et al~\cite{mare2014zebra} \changeMika{describe} a framework for \zebra and implemented \changeMika{some} individual pieces. \changeMika{However, this was not} a complete system. Therefore, we needed to build an end-to-end system from scratch to evaluate our conjecture about opportunistic attacks. Our goal was to make this system as close to the one in \cite{mare2014zebra} as possible. We now describe our system and how we evaluated its performance.

\subsection{Design and Implementation}
\label{sec:system_design}

\noindent\textbf{Software and Hardware}: We followed the ZEBRA system architecture as described in Figure~\ref{fig:architecture}. Our system consists of two applications: the bracelet runs an Android Wear application and the terminal runs a Java application. Interaction Classifier is implemented in Matlab. Communicator modules in both applications orchestrate communication over Bluetooth to synchronize clocks between them and to transfer bracelet measurements to the terminal. The rest of the terminal software consists of the ``\zebra Engine'' (shaded rectangle) with the functionality described in Section~\ref{sec:background}. The bracelet and terminal synchronize their clocks during connection setup. For our experiments, we used a widely available smartwatch (4GB LG G Watch R with a 1.2 GHz CPU and 512MB RAM) with accelerometer/gyroscope as the bracelet and standard PCs as terminals. 

\vspace{1mm}
\noindent\textbf{Parameter Choices}: Mare et al~\cite{mare2014zebra} do not fully describe the parameters used in their implementation of \zebra components. Wherever available, we used the exact parameters provided in \cite{mare2014zebra} \cite{mare}. For the rest, we strived to choose reasonable values. A full list of parameters and rationales for choosing their values appears in Appendix~\ref{app:parameteres}.

\vspace{1mm}
\noindent\textbf{Classifier}: We use the Random
Forest~\cite{breiman2001random} classifier. 
Again, as \cite{mare2014zebra} did not include all details on how their classifier was trained and tuned, we made parameter choices that gave the best results. Our forest consisted of 100 weak-learners. Each split in a tree considered $sqrt(n)$ features, where $n=24$ was the total number of features, and the trees were allowed to fully grow. In addition, classes were weighted to account for any imbalances in the training dataset (described below in Section~\ref{sec:victim-data}).
We adopt the same set of features used in \cite{mare2014zebra}, and extract them for both accelerometer and gyroscope segments. A full list appears in Appendix~\ref{app:features}.

\noindent\textbf{Differences}: Despite our efforts to keep our system similar to that in~\cite{mare2014zebra}, there are some differences. First, we wanted to use commercially widely available general-purpose smartwatches as bracelets. They tend to be less well-equipped compared to the high-end Shimmer Research bracelet used in~\cite{mare2014zebra}. Our smartwatch has a maximum sampling rate of around 200 Hz, whereas the Shimmer bracelet had a sampling frequency of 500 Hz. We discuss the implications of this difference in \changeMika{Section~\ref{subsec:implementation_differences}}.

In addition, \cite{mare2014zebra} mentions a rate of 21 interactions in a 6s period (3.5 interactions per second). However, in our measurements, users filling standard web forms averaged around 1.5 interactions per second. Their typing interactions were slightly less than 1s long on average and MKKM interactions typically spanned 1-1.5s. With our chosen parameters we could produce a rate of 3.5 interactions per second only in sessions involving hectic activity -- switching extremely rapidly between a few key presses and mouse scrolls. Such a high rate could not be sustained in realistic PC usage.


\subsection{Data Collection}
\label{sec:victim-data}

In our study, we recruited 20 participants to serve as users (victims) of the system.
They were mostly students recruited by word of mouth (ages 20--35, 15 males; 5
females, \changeMika{all right-handed)}. 
Participation was voluntary, based on explicit consent. 
The study included both dexterous typists and less-experienced ones.
Initially, we
told the participants that the purpose of the study was to collect information
on how they typically use a PC. At the end of the study, we explained the actual nature of the experiment. The members of our research groups played the role of the adversary \attacker, \changeOtto{compared to the untrained users in~\cite{mare2014zebra}}.
No feedback was given to \attacker whether a given attack attempt was successful or not.

Experiments were conducted in a realistic office setting (with several
other people working at other nearby desks). During a session, a participant did four
10-minute tasks filling a web form, \changeMika{in a similar setting} as in
\cite{mare2014zebra}. From each task, two sets of user data were collected
simultaneously: accelerometer and gyroscope measurements from the user's
bracelet and the actual interaction sequence extracted by Interaction Extractor
on the terminal.  An attacker \attacker assigned to a participant \victim
conducted each of the four types of attack scenarios from
\ref{subsec:scenarios} in turn. In the first three scenarios, \attacker had direct visual
access to \victim. In the fourth scenario, we placed a narrow shoulder-high partition
between \victim and \attacker so that \attacker can hear but not see \victim.
The 20 sessions thus resulted in a total of 80 samples, with each sample
consisting of three traces: bracelet data of the user, actual interaction
sequence of the user, and the actual interaction sequence of the attacker. All
traces within a sample were synchronized. No other information (e.g., the
content of what the participant typed in) was recorded. Participants were told
what data was collected.

The data collection and the study followed IRB procedures at our institutions.
The data we collected has very little personal information. It is
conceivable that the interaction sequences or bracelet data could potentially
be used to link a participant in our study to similar data from the same
participant elsewhere. For this reason, we cannot \changeAsokan{make our datasets 
 public,} but will make them available to other researchers for research use.

\subsection{Performance Evaluation}
\label{sec:performance_evaluation}

\textbf{Usability:} To evaluate {usability}, we follow the same approach as in \cite{mare2014zebra} to compute the false negative rate (FNR) as the fraction of windows in which Authenticator comparing the actual and predicted interaction sequences from the same user incorrectly outputs ``different user.''
We 
employ the leave-one-user-out cross-validation approach: for each session, 
we train a random forest classifier using the 76 samples of bracelet data from all the other 19 sessions. We then use the four samples from the current session to test the classifier. We thus train 20 different classifiers, and report results aggregating classification of 80 samples in all.

\ifllncs
\begin{figure}
\centering
\begin{subfigure}{.5\textwidth}
  \centering
  \includegraphics[width=1\textwidth]{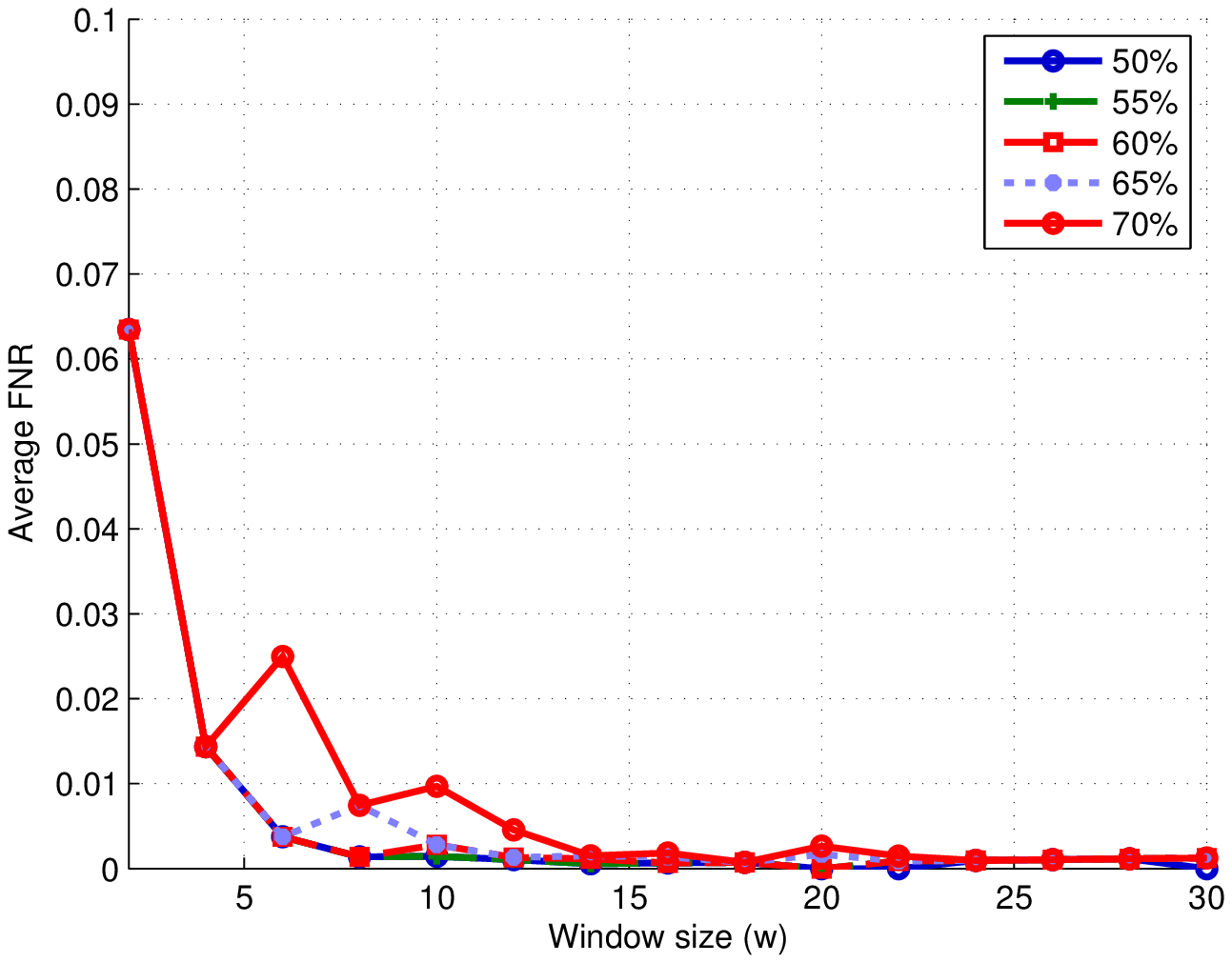}
  \caption{}
  \label{fig:victim_fnr}
\end{subfigure}%
\begin{subfigure}{.5\textwidth}
  \centering
  \includegraphics[width=1\textwidth]{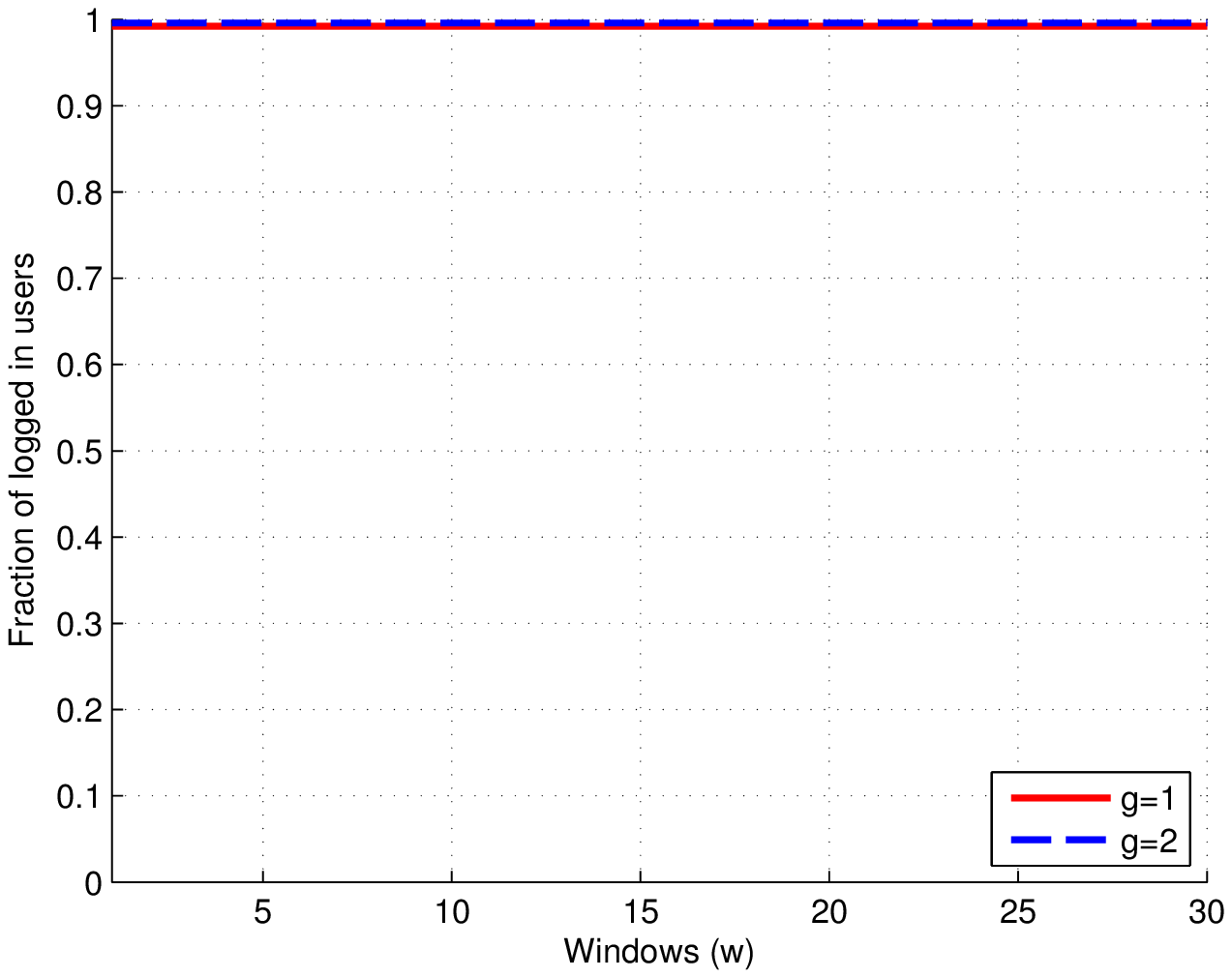}
  \caption{}
  \label{fig:victim_time}
\end{subfigure}
\caption{Performance for legitimate users: \textbf{(a)} Average FNR for different threshold ($m$)values. \textbf{(b)} Fraction of users remaining logged in after ($n$) authentication windows (with $w=20, m=60\%)$, for different grace periods ($g$).}
\end{figure}

\else

\begin{figure}[bh!]
	\centering
	\begin{subfigure}[b]{1\linewidth}
	\centering
		\includegraphics[width=0.7\linewidth]{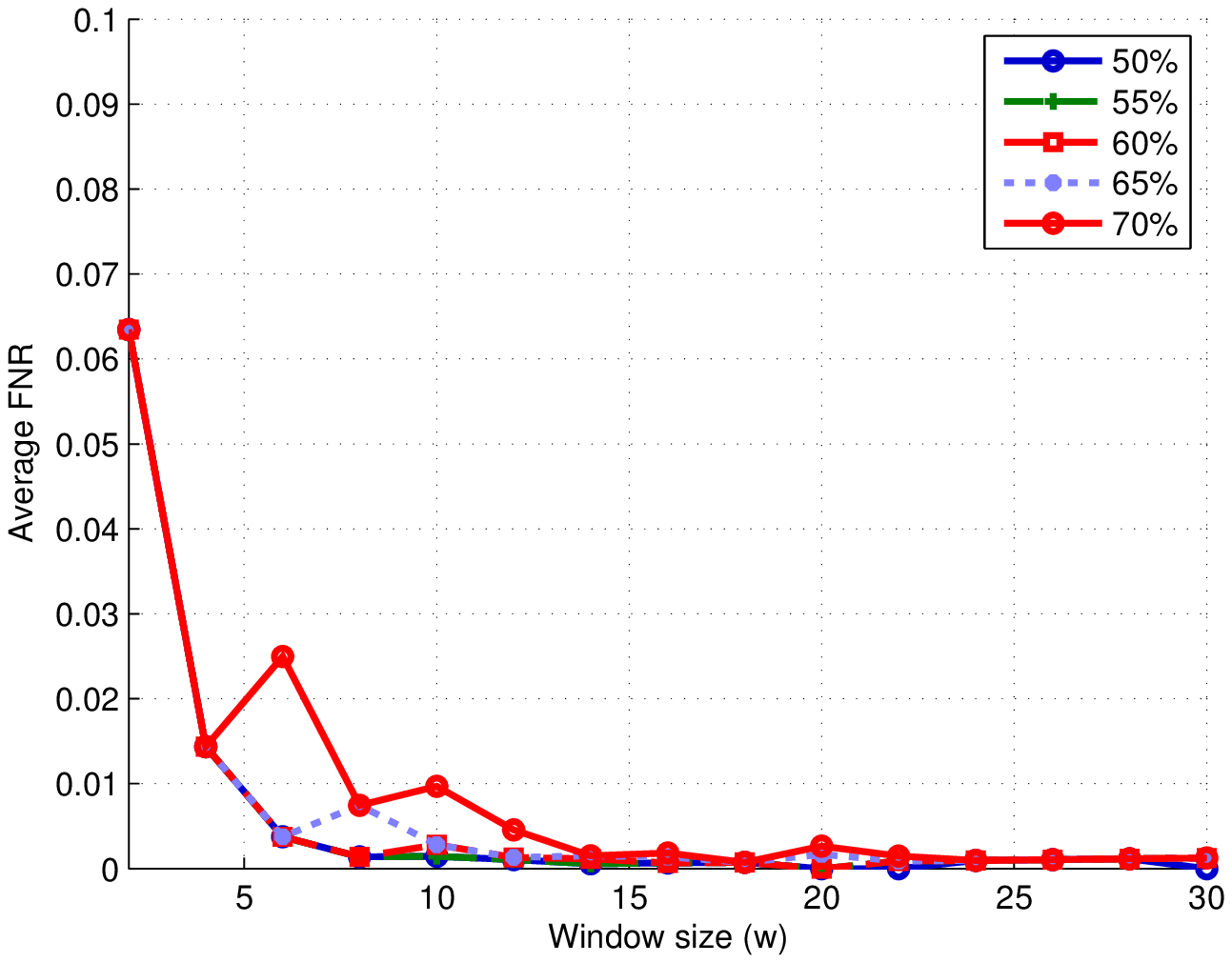}
		\caption{Average FNR vs. window size ($w$) for different threshold ($m$) values. \changeMika{Fraction of windows that are incorrectly classified as mismatching.}}
		\label{fig:victim_fnr}
	\end{subfigure}

	\begin{subfigure}[b]{1\linewidth}
	\centering
		\includegraphics[width=0.7\linewidth]{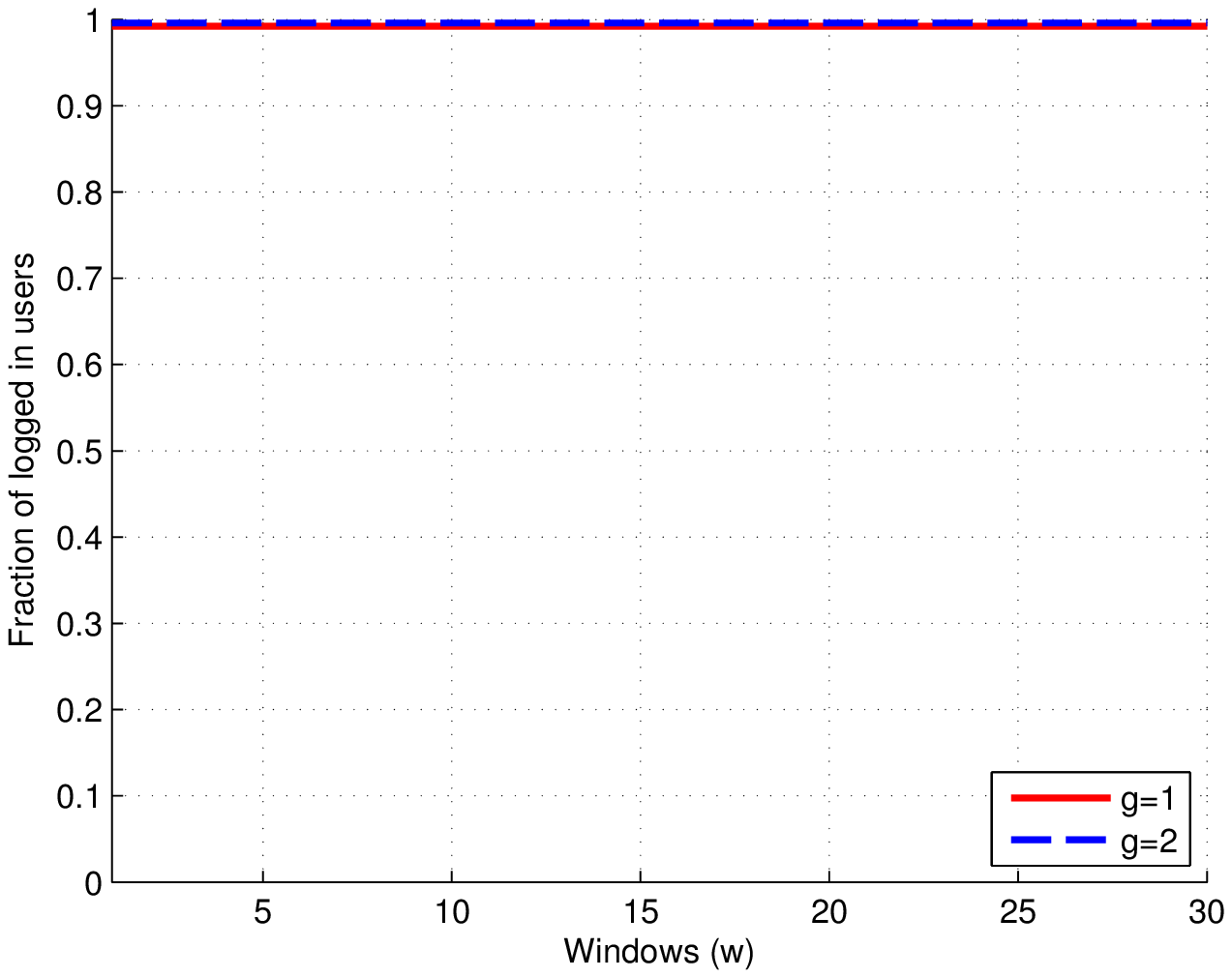}
		\caption{Fraction of users remaining logged in after ($n$) authentication windows (with $w=20, m=60\%)$, for different grace periods ($g$).}
		\label{fig:victim_time}
	\end{subfigure}
	\caption{Performance for legitimate users}
\end{figure}

\fi


Figure~\ref{fig:victim_fnr} shows how different window size ($w$) and matching threshold ($m$) values affect average FNR. 
As can be seen, FNR is very low for our system. The original \zebra paper \cite{mare2014zebra} reports FNRs in the range of 0-16\% whereas in our system the FNRs are 0-6\%, and below 1\% for window sizes above 10. 

We also estimated the length of time (in terms of the number of windows) for which a legitimate user remained logged in. For this, we fix $w=20$ and $m=60\%$ as in \cite{mare2014zebra}. On average, a window was 13 seconds long. The low FNRs result in no legitimate users getting logged out in any of the 10 minute samples. Figure~\ref{fig:victim_time} depicts this by plotting the fraction of users still logged in after a given number of authentication windows. 
The situation is the same when allowing one additional failed authentication window before logging a user out ($g=2$), or when directly logging the user out after the first failed window ($g=1$). This also seems in line with the results reported in \cite{mare2014zebra}, where one legitimate user was logged out when using a stricter grace period ($g=1$). 


Table~\ref{fig:cm} presents the confusion matrix for the classification performance of our Interaction Classifier. It combines data for all 80 (20 x 4) classifications. 
\changeMika{It shows that our system is very} \changeAsokan{good at} \changeMika{recognizing events accurately.}
For example, for the \textit{typing} events, we obtain a precision of 96.9\% ($15753/16252$) and a recall of 96.5\% ($15753/16332$).

\begin{table}[h]
	\centering
	\begin{center}
	\caption{Confusion matrix for 80 legitimate user samples.}
	\begin{tabular}
	{|c| >{\bfseries}r @{\hspace{0.7em}}|c @{\hspace{0.7em}}|c @{\hspace{0.7em}}|c|}
	\hline
	\multirow{7}{*}{\rotatebox{90}{\parbox{1.1cm}{\bfseries\centering Actual}}} & 
	\multicolumn{3}{c}{\bfseries Predicted} & \\[0.5em]
	\hline
	& & \bfseries Typing & \bfseries Scrolling & \bfseries MKKM  \\[0.5em]
	\cline{2-5}
	& Typing & 15753 & 354 & 225 \\[0.5em]
	\cline{2-5}
	& Scrolling & 271 & 2506 & 2 \\[0.5em]
	\cline{2-5}
	& MKKM & 228 & 71 & 15378 \\[0.5em]
	\hline
	\end{tabular}
	\label{fig:cm}
	\end{center}
\end{table}

\vspace{1mm}
\noindent\textbf{Detection of Innocent Adversaries}:
To estimate the security against an innocent adversary (a different user) who inadvertently starts using an unattended terminal where another user has logged in, we compute the true negative rate (TNR) for ``mismatching'' sequences: where the actual interaction sequence of one sample is compared against the predicted interaction sequence of a \textit{different sample}. With such mismatched sequences, the TNR is the fraction of windows in which the ``wrong'' user is correctly classified as ``different user.''  Recall that data within a sample (and thus the interaction sequences extracted from it) are synchronized. When mismatching samples to compute TNR, we synchronized traces by aligning the starting points of the sequences being compared. 

Figure~\ref{fig:wrongvictim_tnr} shows how different $w$ and $m$ values impact the average TNR (over 20 x 4 classifications) of our system with mismatched traces as input. Especially for thresholds of 60-70\%, a majority of the authentication windows are identified correctly as non-matching. Again, using $w=20$ and $m=60\%$, Figure~\ref{fig:wrongvictim_time} shows the fraction of ``wrong'' users who remain logged in (i.e., incorrectly \textit{not} deauthenticated) after interacting with the terminal for a given number of windows. 

\ifllncs

\begin{figure}
\centering
\begin{subfigure}{.5\textwidth}
  \centering
  \includegraphics[width=1\textwidth]{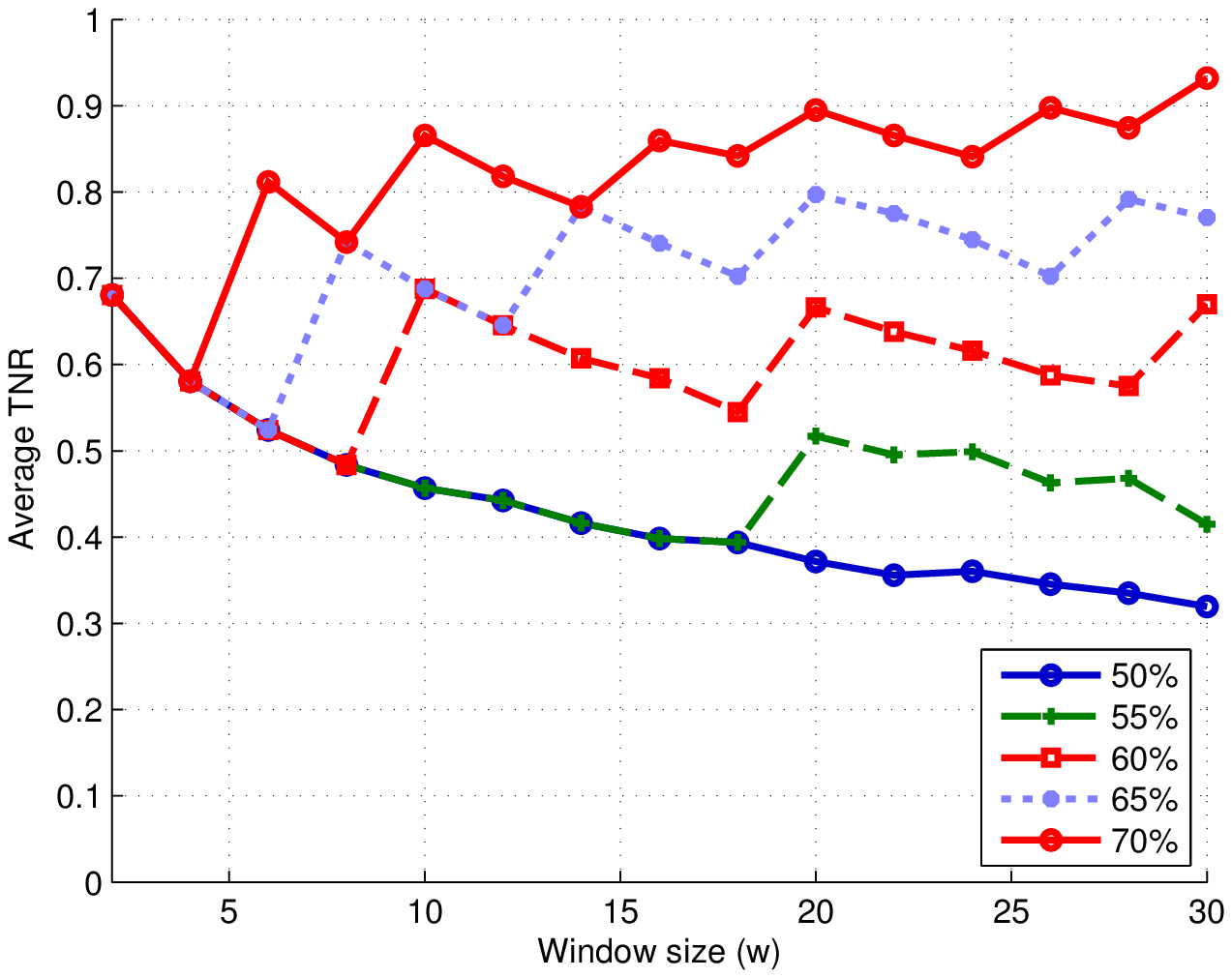}
  \caption{}
  \label{fig:wrongvictim_tnr}
\end{subfigure}%
\begin{subfigure}{.5\textwidth}
  \centering
  \includegraphics[width=1\textwidth]{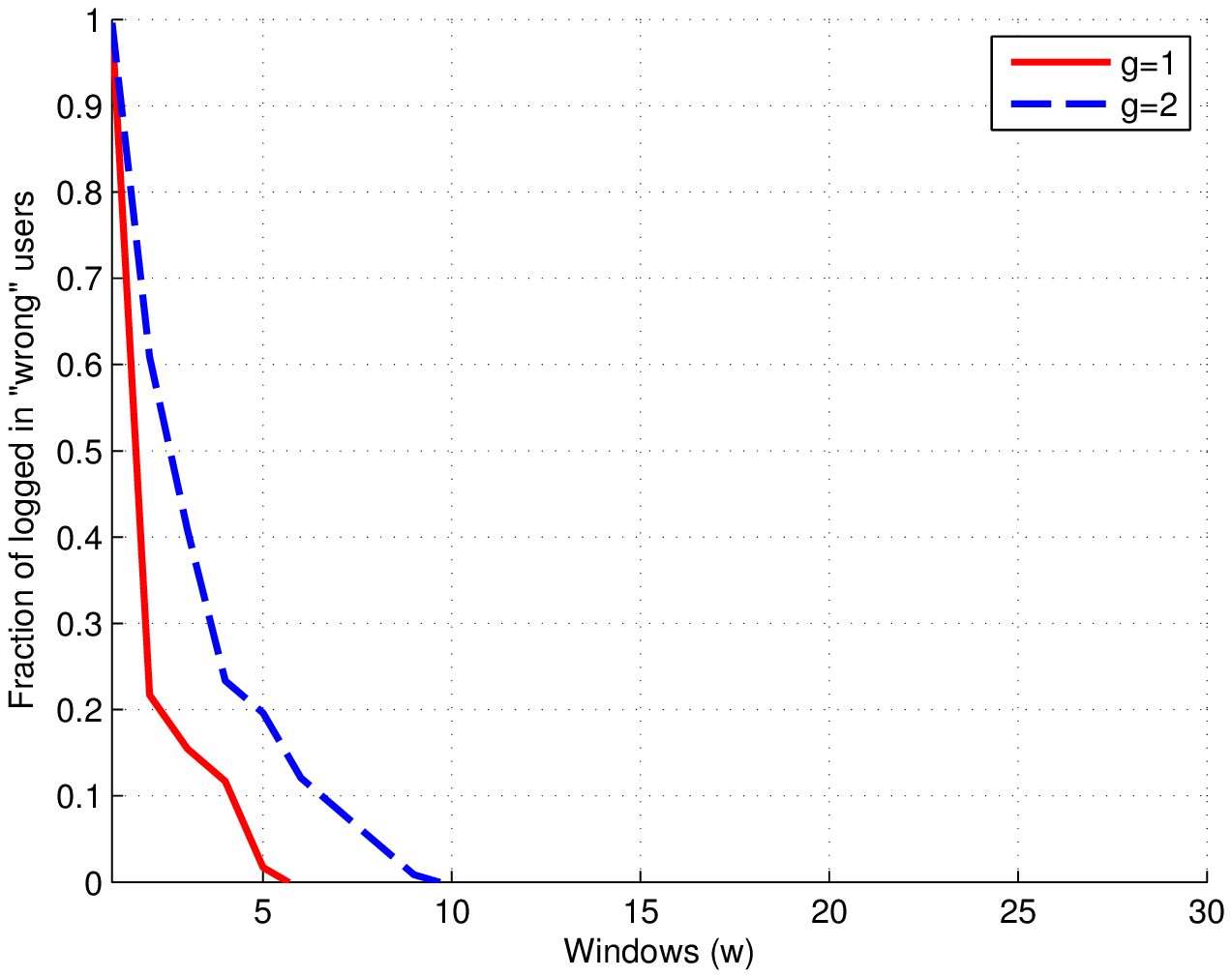}
  \caption{}
  \label{fig:wrongvictim_time}
\end{subfigure}
\caption{Performance for ``wrong'' (mismatched) users: \textbf{(a)} Average TNR for different threshold ($m$) values. \textbf{(b)} Fraction of users remaining logged in after ($n$) authentication windows (with $w=20, m=60\%)$, for different grace periods ($g$).}
\end{figure}

\else

\begin{figure}[t!]
	\centering
	\begin{subfigure}[b]{1\linewidth}
	\centering
		\includegraphics[width=0.7\linewidth]{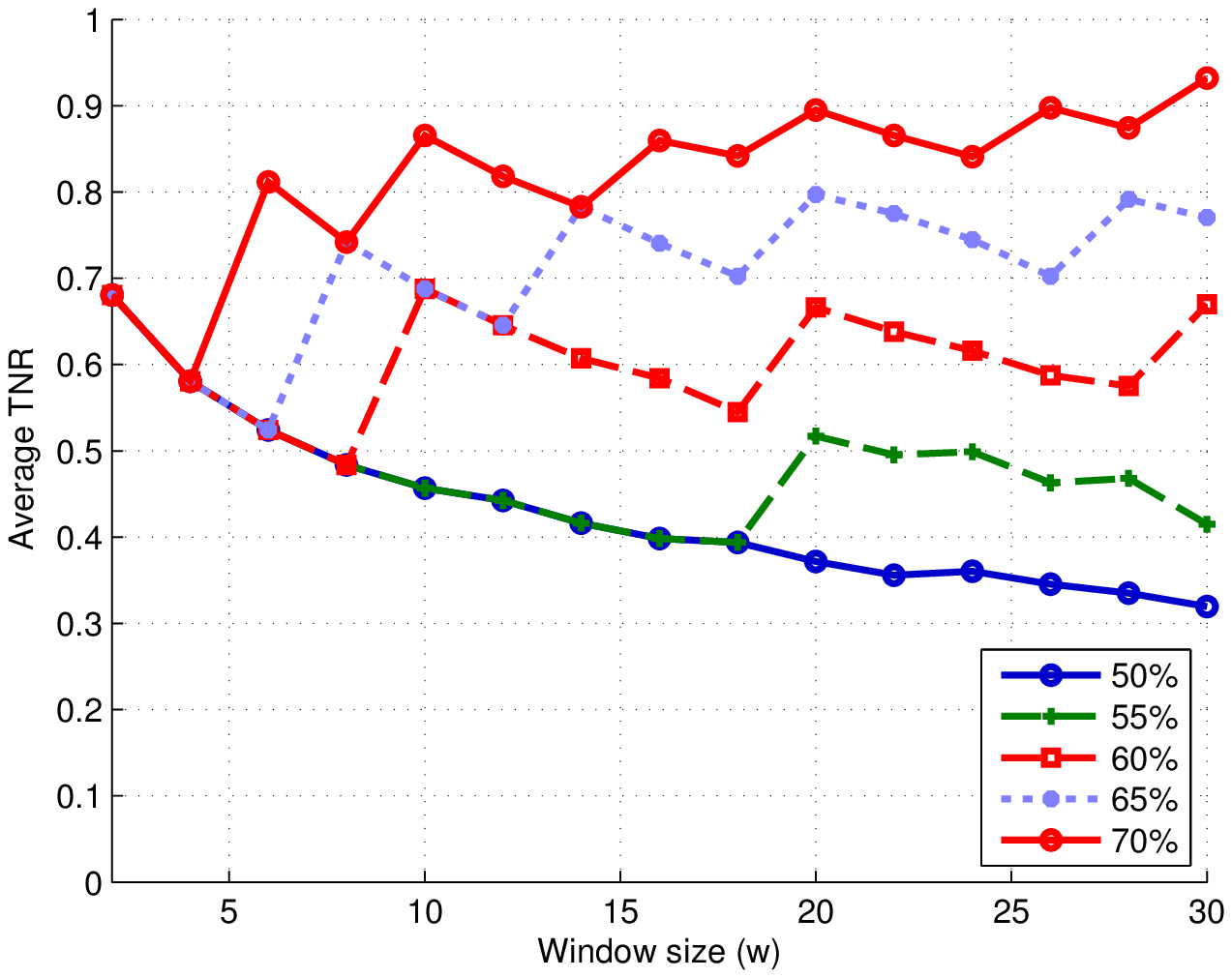}
		\caption{Average TNR for different threshold ($m$) values. \changeMika{Fractions of windows that correctly identify a wrong user.}}
		\label{fig:wrongvictim_tnr}
	\end{subfigure}

	\begin{subfigure}[b]{1\linewidth}
	\centering
		\includegraphics[width=0.7\linewidth]{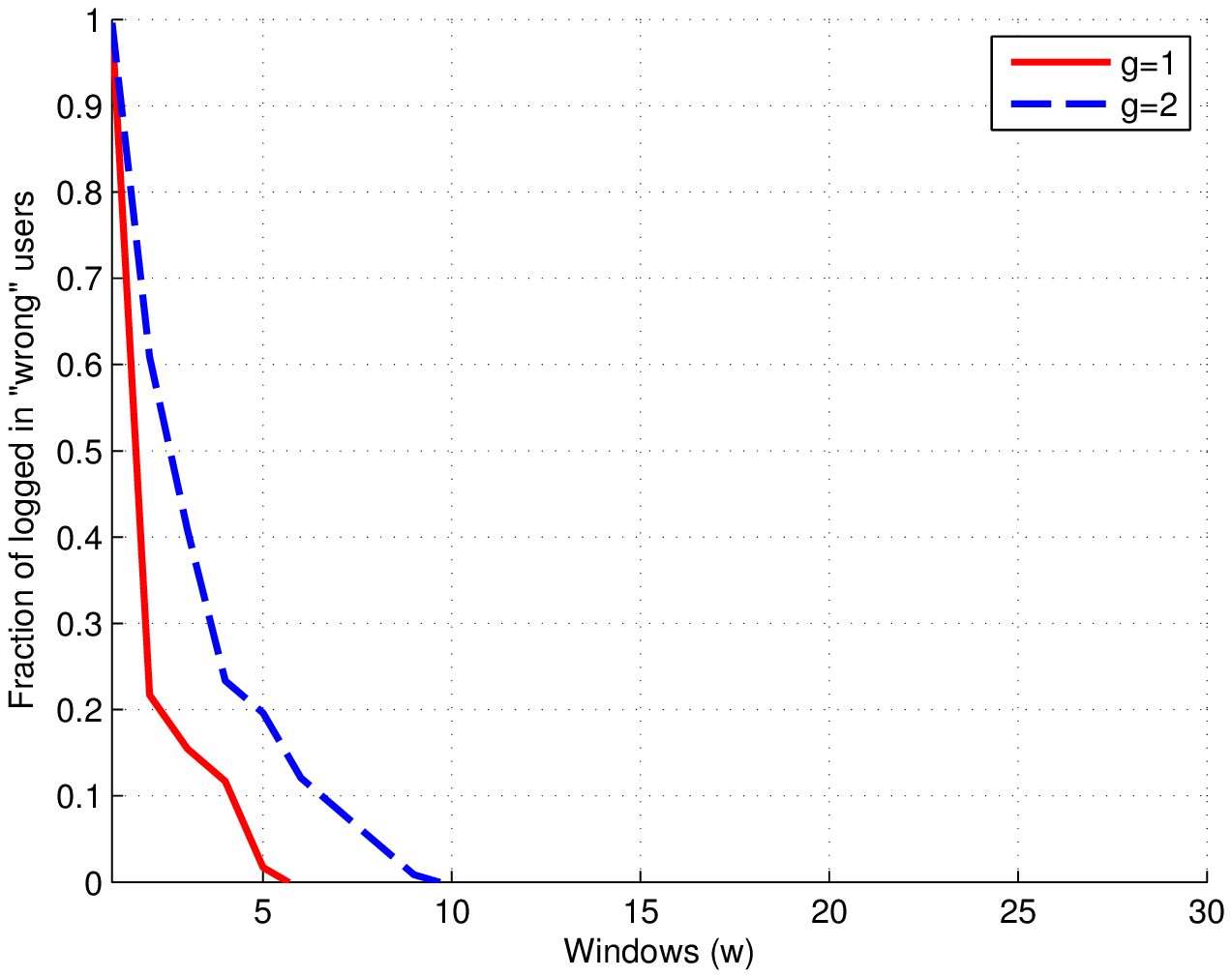}
		\caption{Fraction of users remaining logged in after ($n$) authentication windows (with $w=20, m=60\%)$, for different grace periods ($g$).}
		\label{fig:wrongvictim_time}
	\end{subfigure}
	\caption{Performance for ``wrong'' (mismatched) users. \changeMika{Simulated accidental usage of the terminal.}}
\vskip-0.5cm
\end{figure}

\fi

When the legitimate user is also interacting with a terminal, it can be expected that a non-zero fraction of actual interactions by the ``wrong user'' will accidentally match the predicted interactions by the legitimate user. As such \zebra Authenticator will accept (output 1) for a fraction of authentication windows. However, as can be seen from the fraction of logged in users in Figure~\ref{fig:wrongvictim_time}, a majority of users will quickly get logged out as any such accidental matches are not sufficient to keep the user logged in for an extended period of time. Using a strict grace period ($g=1$), 78\% of wrong users are logged out after the first authentication window and all but one after 5 windows. For $g=2$, 80\% of wrong users are logged out after 5 windows, and all by window 10. 

To further evaluate the resilience of our system, we investigated the impact on FNR if the predicted and actual interaction sequences are desynchronized. We shifted the actual interaction sequence in each sample forward in time to simulate the time delay incurred, for example, when an attacker mimics his victim. Delays of 200 ms increase the Negative Rates (NR) from the 0-6\% (presented in Figure~\ref{fig:victim_fnr}) to 1-20\%, resulting in 5-10\% of legitimate users getting logged out. Further increasing the delay to 500 ms increases the NR to 25-70\% causing a majority of users to be logged out within 2-4 authentication windows. Thus, despite its low FNRs for legitimate users, our system is robust because it is sensitive to delays introduced in mimicking user interactions.

\vspace{1mm}
\noindent\textbf{Summary}:
We therefore conclude that our \changeMika{end-to-end} system is functionally comparable to that of \cite{mare2014zebra}. Legitimate users remain logged-in at a very high rate, whereas the majority of wrong users are quickly logged out. 
Our system achieves lower FNR for legitimate users compared to \cite{mare2014zebra}, which is good for usability but may also be caused if the system is too permissive. However, our experiments with mismatched and desynchronized traces show marked increases in 
\changeMika{FNR} 
suggesting that our system is not overly permissive.


%% file: chapters/6results.tex
\section{Malicious Adversaries}
\label{sec:results}

Having shown that our \changeMika{end-to-end} system is resilient against innocent adversaries, we now consider its security against malicious adversaries who attempt to intentionally mimic a victim's interactions. We consider the four types of attack scenarios from Section~\ref{subsec:scenarios}: na\"ive and opportunistic \allactivity attacks, and two variants of opportunistic \KBactivity attacks.

In all four cases, we use data from the 20 user sessions. As before, we use the leave-one-user-out approach: for a given session, we train Interaction Classifier using the bracelet traces from the 76 samples from the remaining 19 sessions. For each type of attack, we then apply the classifier for the corresponding trace in the current sample. Thus, the results for each attack scenario is the aggregated result of 20 classifications.

\vspace{1mm}
\noindent
\textbf{Na\"ive \allactivity}:
Figure~\ref{fig:all_full_fpr} presents the average False Positive Rate (FPR) for threshold values ($m$) between 50\% and 70\%, and for window sizes ($w$) in the 5-30 range. The FPR represents the fraction of authentication windows in which the attacker is mistaken for the victim, i.e., a large enough fraction of interactions are evaluated as matching. The FPR values range from 50-80\% with a lenient threshold of 50\%, and from 15-35\% with a strict threshold of 70\%. For example, with $m=70\%$ and $w=20$, less than one fifth of the attackers' authentication windows are correct.

\ifllncs

\captionsetup[table]{font=small,skip=0pt}

\begin{figure}
	\vspace{-5mm}
\centering
\begin{subfigure}{.5\textwidth}
  \centering
  \includegraphics[width=\textwidth]{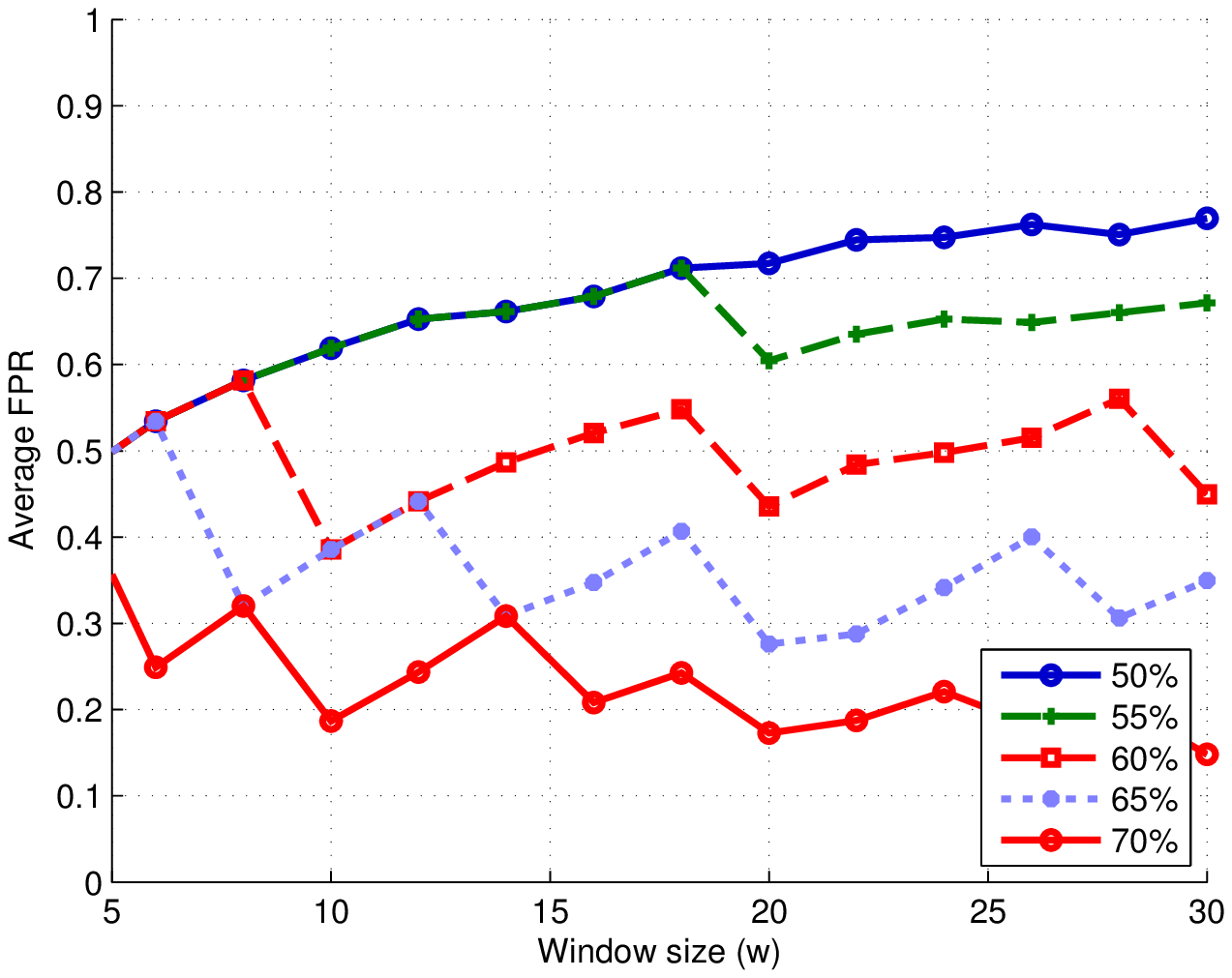}
   \caption{}
  \label{fig:all_full_fpr}
\end{subfigure}%
\begin{subfigure}{.5\textwidth}
  \centering
  \includegraphics[width=\textwidth]{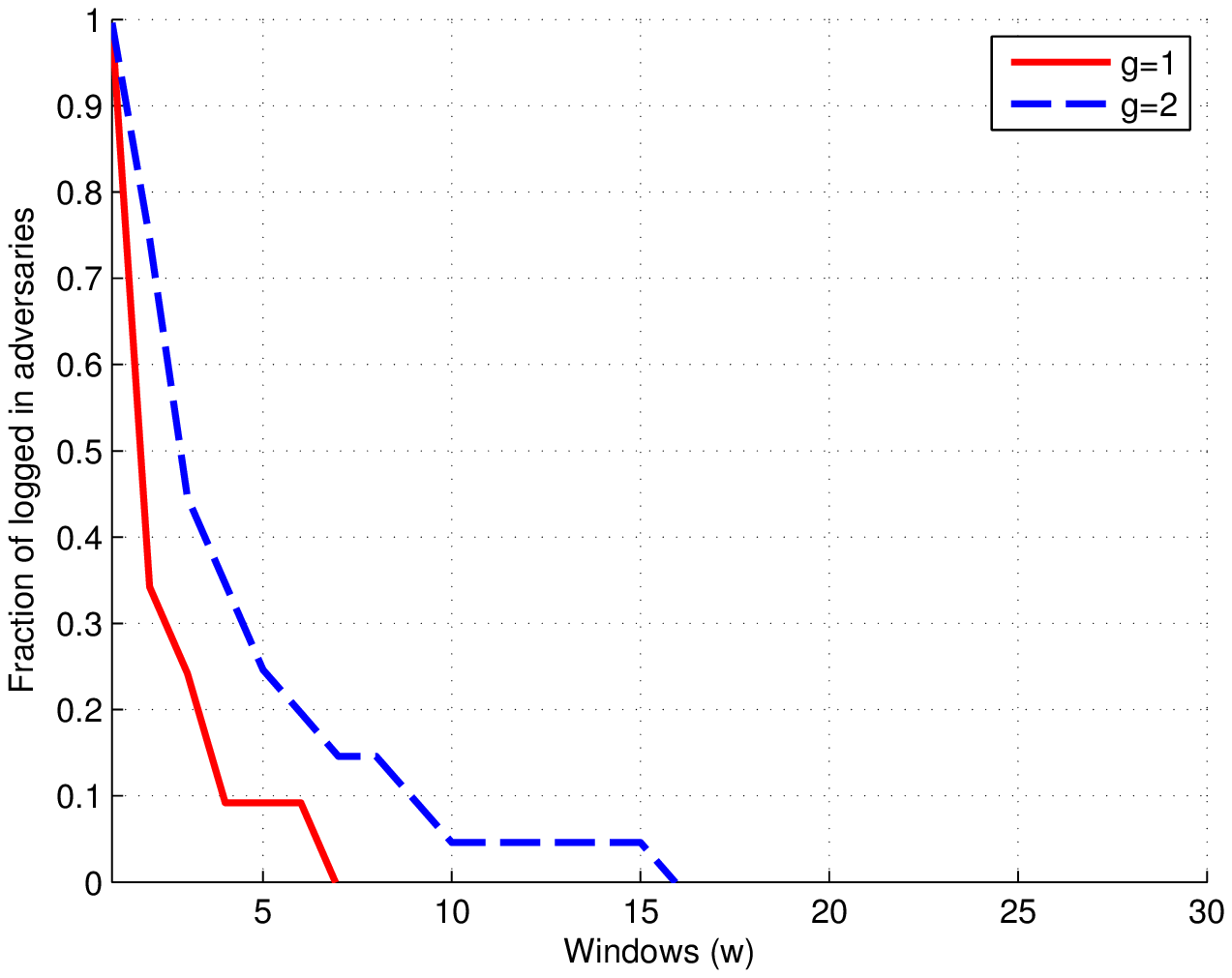}
  \caption{}
  \label{fig:all_full_time}
\end{subfigure}
\caption{\textbf{Na\"ive \allactivity} attacker: \textbf{(a)} Average
  FPR for different threshold ($m$) values. \textbf{(b)} Fraction of
  attackers remaining logged in after ($n$) authentication windows (with $w=20, m=60\%)$, for different grace periods ($g$).}
	\vspace{-5mm}
\end{figure}

\else

\begin{figure}[t!]
	\centering
	\begin{subfigure}[b]{\figwidth}
	\centering
		\includegraphics[width=0.7\figwidth]{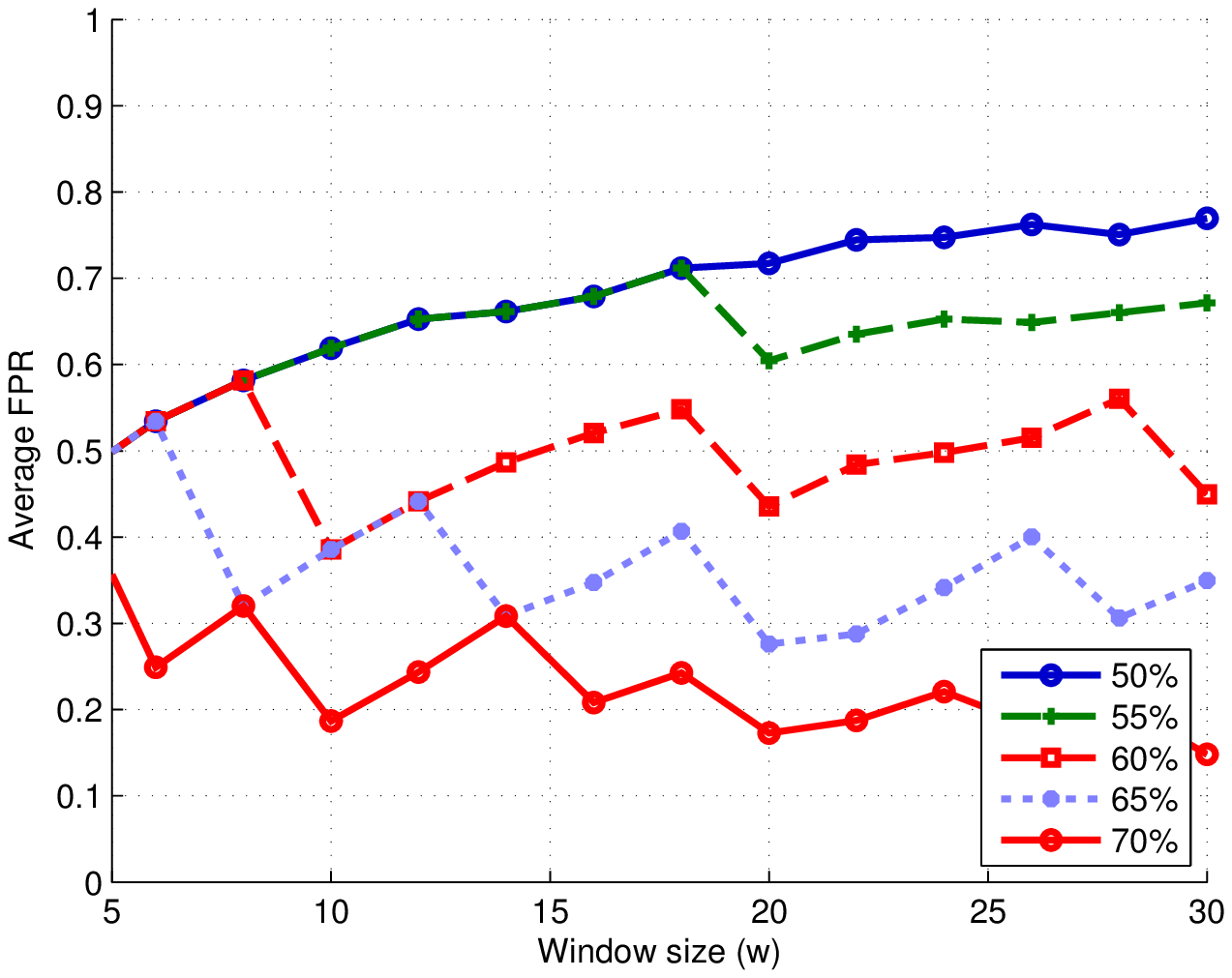}
		\caption{Average FPR for different threshold ($m$) values. \changeMika{Fraction of attacker windows that are classified as matching with the bracelet.}}
		\label{fig:all_full_fpr}
	\end{subfigure}

	\begin{subfigure}[!b]{\figwidth}
	\centering
		\includegraphics[width=0.7\figwidth]{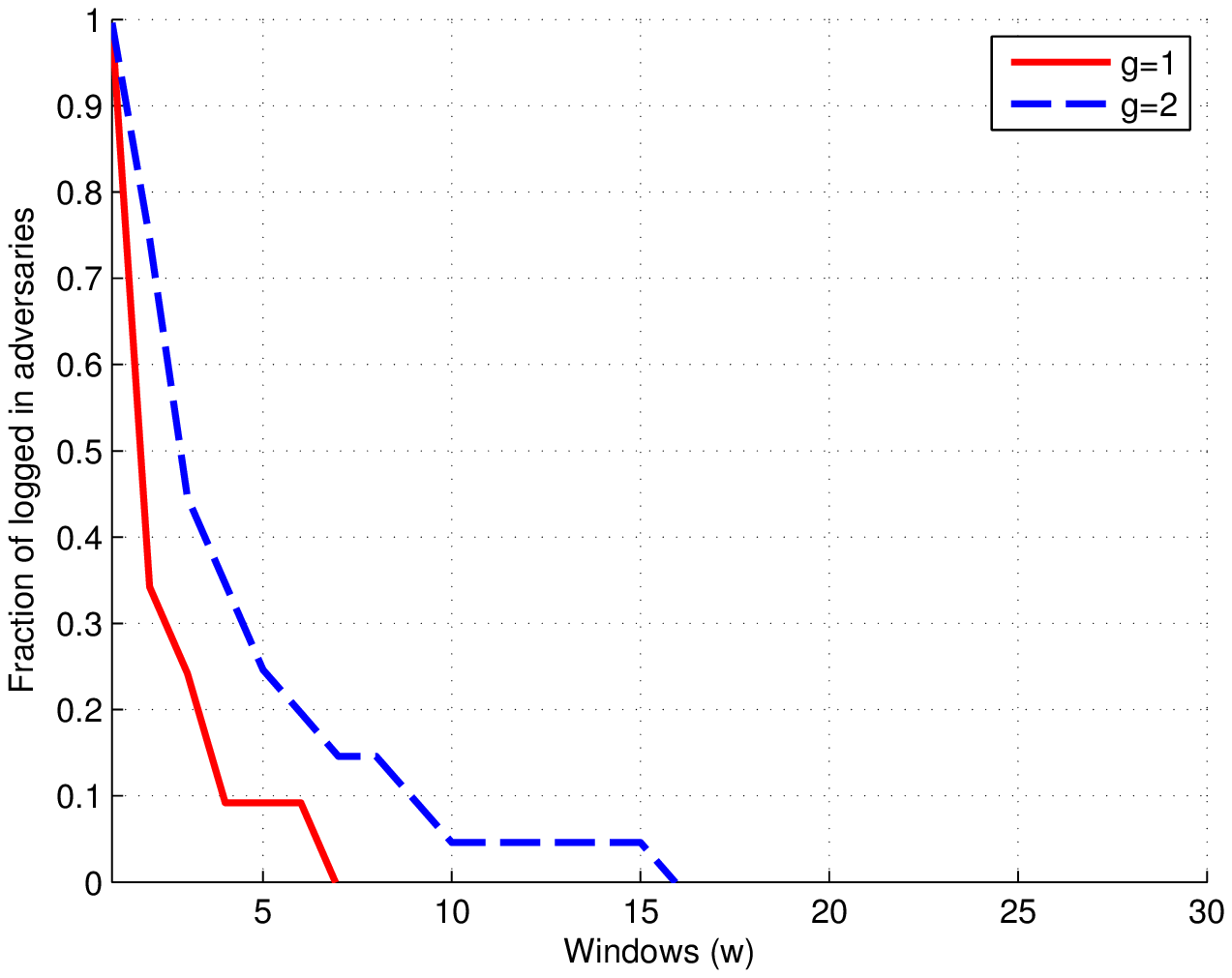}
		\caption{Fraction of attackers remaining logged in
                  after ($n$) authentication windows, for different grace periods ($g$).}
		\label{fig:all_full_time}
	\end{subfigure}
	\caption{\changeMika{Results for \textbf{na\"ive \allactivity} attackers. Na\"ive \allactivity attackers try to replicate all mouse-hand movements.}}
\end{figure}

\fi

We choose the same threshold and window size as previously described ($m=60\%$ and $w=20$), and determine the fraction of logged in users as a function of the number of authentication windows. This represents how long the attackers successfully remain logged in. Figure~\ref{fig:all_full_time} depicts this fraction for $g=1,2$ . The FPR of 43\% from Figure~\ref{fig:all_full_fpr} translates to all users eventually being logged out. With a strict grace period ($g=1$) all attackers are logged out by the seventh authentication window, whereas with $g=2$ one attacker remains logged in until window 16 \changeMika{(all others fail at window 10 at the latest)}. The victim in this one case had very slow interactions, which made them easier to mimic. \changeMika{The corresponding number of windows in the \zebra paper \cite{mare2014zebra} were 2 and 4.}

The na\"ive \allactivity attacker is comparable to the attacker modeled in \cite{mare2014zebra}. However, the performance of our system against such an attacker (as summarized in Figures~\ref{fig:all_full_fpr} and \ref{fig:all_full_time}) is more lenient than the corresponding figures reported in \cite{mare2014zebra}. Nevertheless, we can use the results for the na\"ive \allactivity attacker as a baseline to compare against more sophisticated or smart attacker strategies we study next.

\ifllncs

\begin{figure}[!b]
	\vspace{-5mm}
\centering
\begin{subfigure}{.5\textwidth}
  \centering
  \includegraphics[width=1\textwidth]{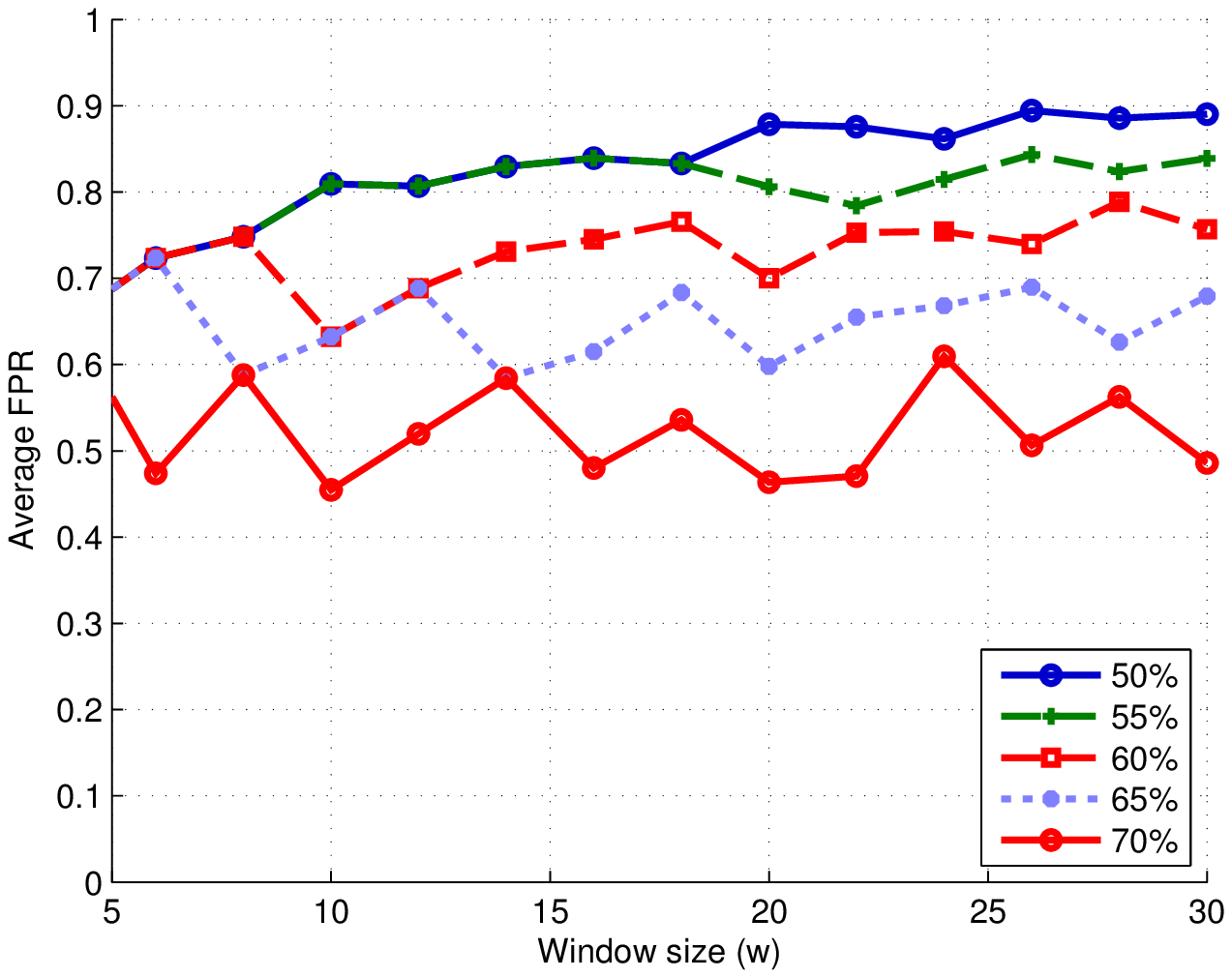}
  \caption{}
  \label{fig:kb_smart_fpr}
\end{subfigure}%
\begin{subfigure}{.5\textwidth}
  \centering
  \includegraphics[width=1\textwidth]{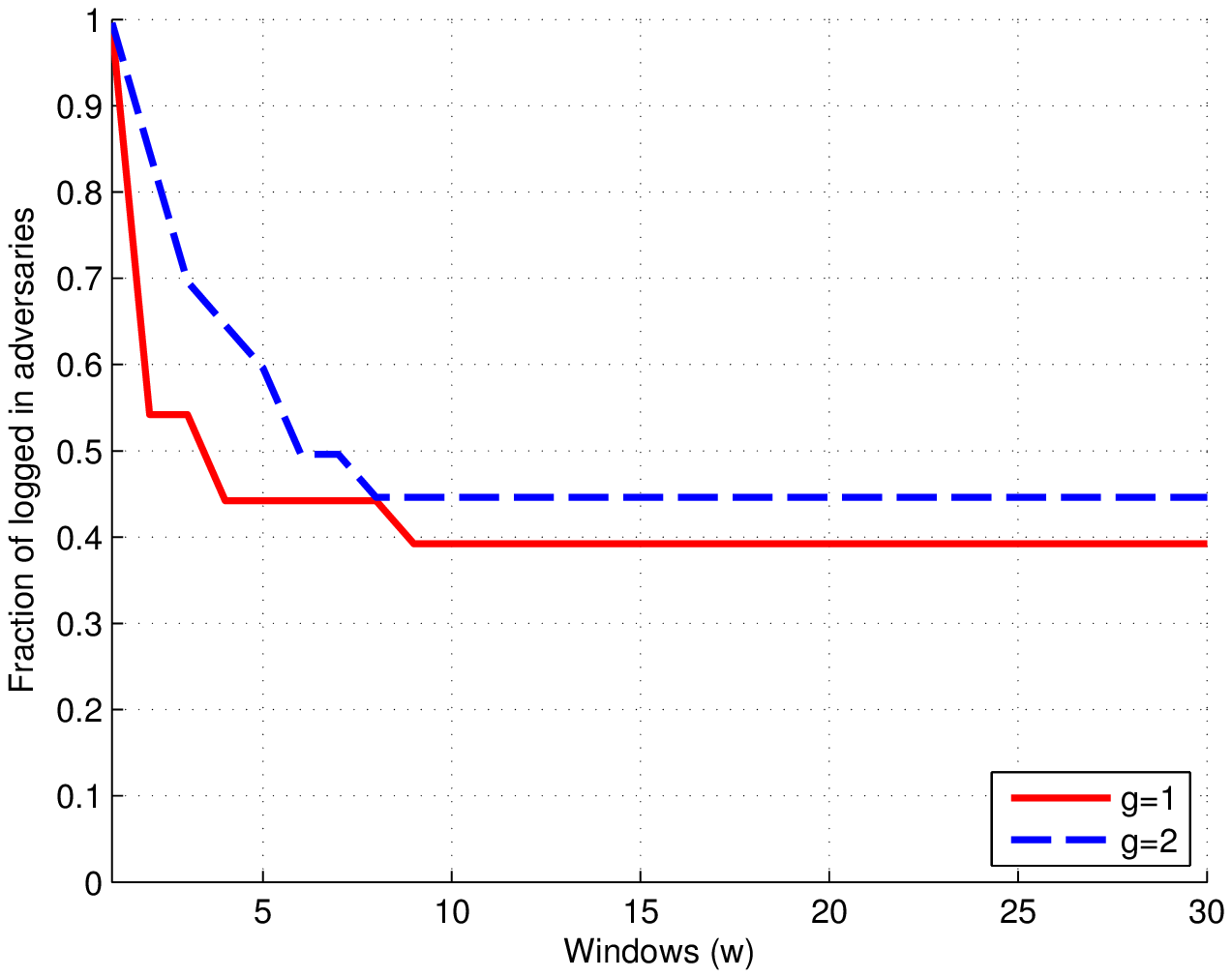}
  \caption{}
  \label{fig:kb_smart_time}
\end{subfigure}
\caption{\textbf{Opportunistic \KBactivity} attacker: \textbf{(a)}
  Average FPR for different threshold ($m$) values. \textbf{(b)}
  Fraction of attackers remaining logged in after ($n$) authentication
  windows (with $w=20, m=60\%)$, for different grace periods ($g$).}
\end{figure}
\else
\begin{figure}[t!]
	\centering
	\begin{subfigure}[b]{\figwidth}
	\centering
		\includegraphics[width=0.7\figwidth]{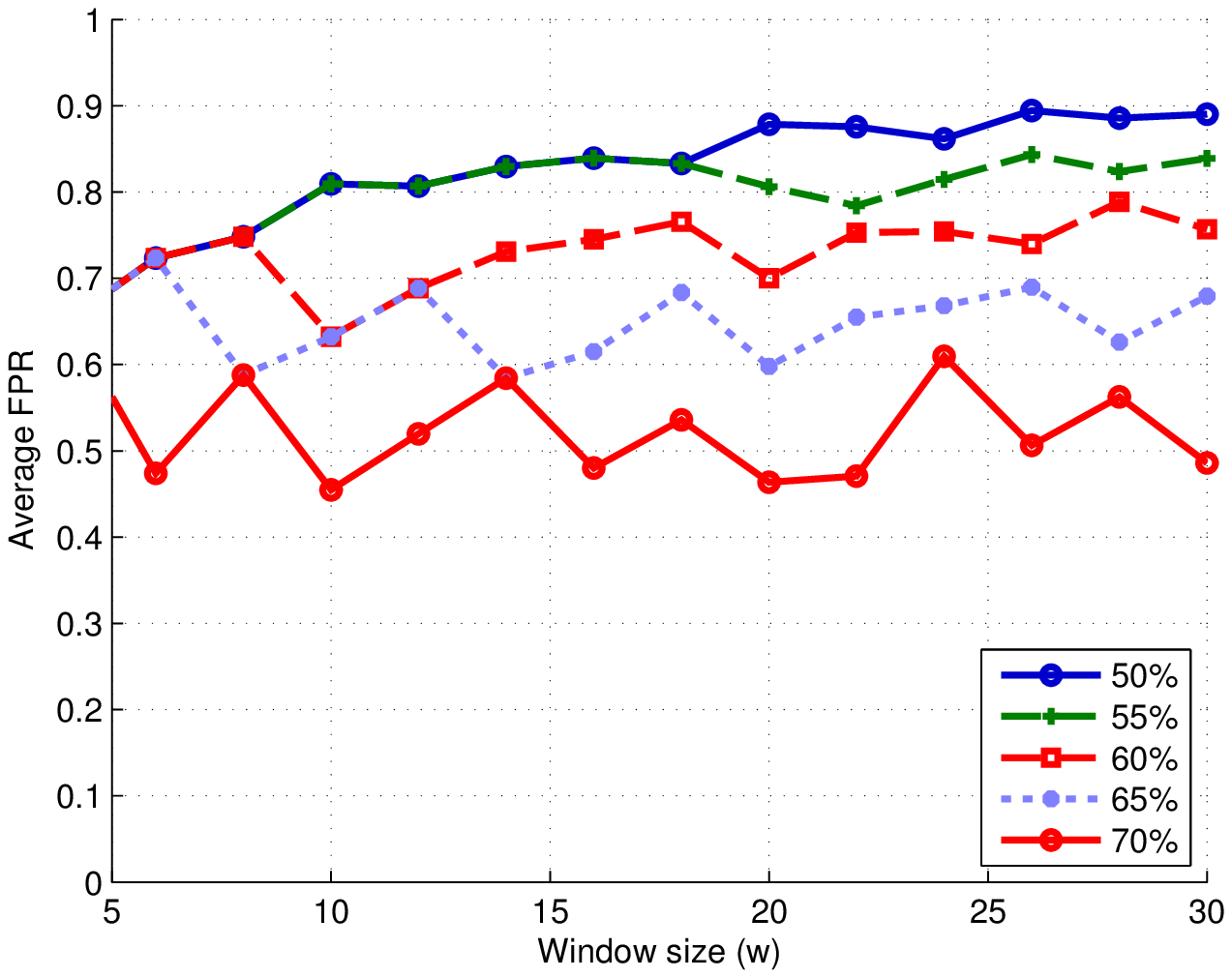}
		\caption{Average FPR for different threshold ($m$) values. \changeMika{Fraction of attacker windows that are classified as matching with the bracelet.}}
		\label{fig:kb_smart_fpr}
	\end{subfigure}

	\begin{subfigure}[b]{\figwidth}
	\centering
		\includegraphics[width=0.7\figwidth]{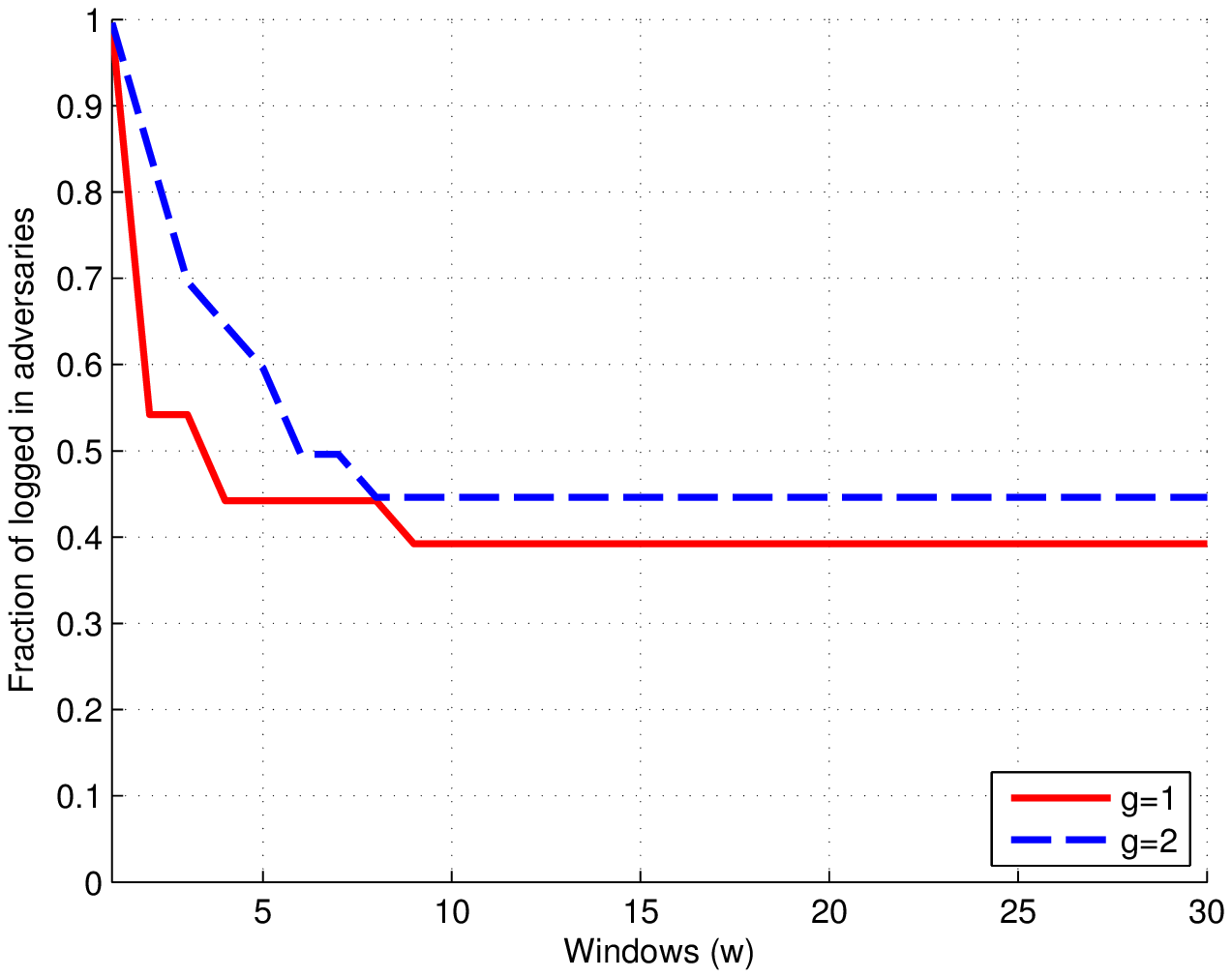}
		\caption{Fraction of attackers remaining logged in
                  after ($n$) authentication windows for different grace periods ($g$).}
		\label{fig:kb_smart_time}
	\end{subfigure}
	\caption{\changeMika{Results for \textbf{opportunistic \KBactivity} attackers. Opportunistic \KBactivity attackers choose to replicate only a part of the keyboard movements of the victim.}}
	\vspace{-5mm}
\end{figure}
\fi

\begin{figure}[!t]
  \centering
  \includegraphics[width=0.7\columnwidth]{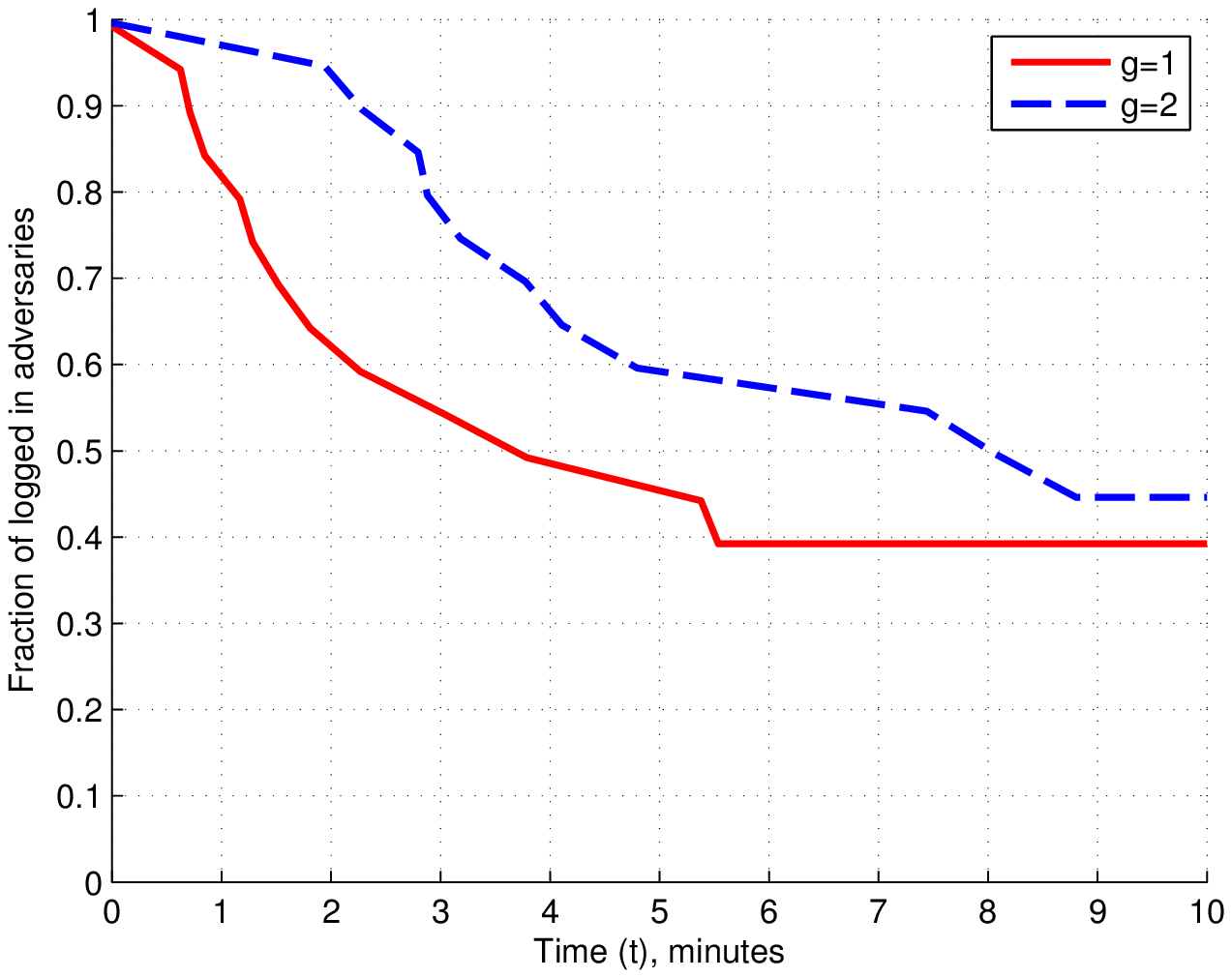}
\caption{\textbf{Opportunistic \KBactivity} attacker: 
  Fraction of attackers remaining logged in after ($t$) minutes
  (with $w=20, m=60\%)$, for different grace periods ($g$).}
  \label{fig:kb_smart_time_sec}
\end{figure}

\vspace{1mm}
\noindent
\textbf{Opportunistic \KBactivity}:
We now consider an attacker who opportunistically mimics only a subset of the typing interactions. Figure~\ref{fig:kb_smart_fpr} presents average FPR for different threshold values and window sizes. The FPRs are now noticeably higher. A threshold of $m=60\%$ and a window size of $w=20$ now produces an FPR of 70\%. Even with a stricter threshold of 70\%, in around half of the windows, attacker interactions are incorrectly evaluated as matching the victim's interactions. \changeMika{In summary, windows are misclassified as correct ones roughly 20 percent points more often with an opportunistic \KBactivity attacker, compared to a na\"ive \allactivity attacker.}

These high FPRs translate to almost half of the attackers remaining successfully logged in for the whole duration of the experiment. Figure~\ref{fig:kb_smart_time} depicts the fraction of logged in attackers as a function of the number of authentication windows, using $g=1,2$. Figure~\ref{fig:kb_smart_time_sec} shows the same information in terms of minutes. In terms of remaining successfully logged in, the advantage of an opportunistic \KBactivity attacker (Figure~\ref{fig:kb_smart_time}) over the na\"ive \allactivity attacker (Figure~\ref{fig:all_full_time}) is statistically significant
(Wilcoxon signed-rank test, $z=-2.928$ and $p=0.003 \ll 0.05$) with medium
effect size ($r=-0.46)$. 
\changeMika{In other words, \KBactivity attackers remain logged in statistically longer than \allactivity attackers.}
Using $g=1$ results in 40\% of the attackers remaining logged in throughout the experiment. A grace period of $g=2$ increases this to 45\%.

Given that an opportunistic \KBactivity attacker can do significantly better than the na\"ive \allactivity attacker, we conclude that the attack scenario used in \cite{mare2014zebra} to demonstrate the security of \zebra is \textit{not the most favorable setting} for the adversary. Also, in our experiments the opportunistic attackers reproduced around 60\% of the victims' typing interactions, reaching typing speeds of 20-40 words/minute. Even at this high typing rate, 40-45\% of attackers were able to successfully evade detection throughout the experiment. A more conservative strategy would naturally increase the attacker success rates closer to 100\%.
To clarify, the number of interactions generated per unit of time is not bound to the typing speed: \zebra concatenates consecutive typing events into a single typing interaction of up to 1s in length. Victims who type slowly may give the attacker more time to mimic. For a malicious attacker, even a short period is enough to cause damage to the system. The attacker can, for example, mount a USB drive and execute a script from the drive in mere seconds.

\vspace{1mm}
\noindent \textbf{Other Attacks:} Having demonstrated that opportunistic \KBactivity attacks are effective, we now consider two variations. First we ask whether the opportunistic approach can be extended successfully to mimicking all types of activities rather than just typing. Figures~\ref{fig:all_smart_fpr} and \ref{fig:all_smart_time}
\ifllncs

\else
\begin{figure}[t!]
	\centering
	\begin{subfigure}[b]{\figwidth}
	\centering
		\includegraphics[width=0.7\figwidth]{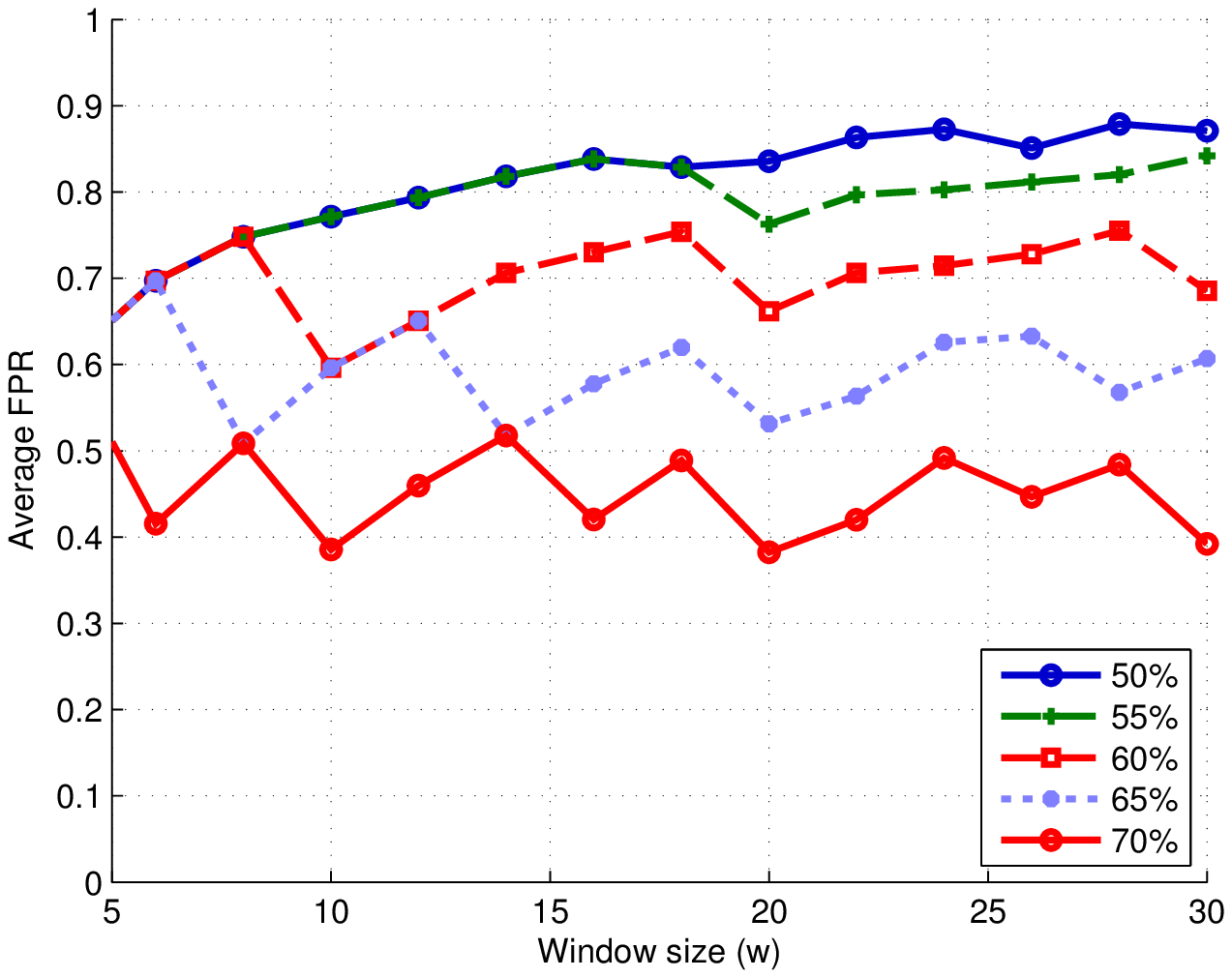}
		\caption{Average FPR for different threshold ($m$) values. \changeMika{Fraction of attacker windows that are classified as matching with the bracelet.}}
		\label{fig:all_smart_fpr}
	\end{subfigure}

	\begin{subfigure}[b]{\figwidth}
	\centering
		\includegraphics[width=0.7\figwidth]{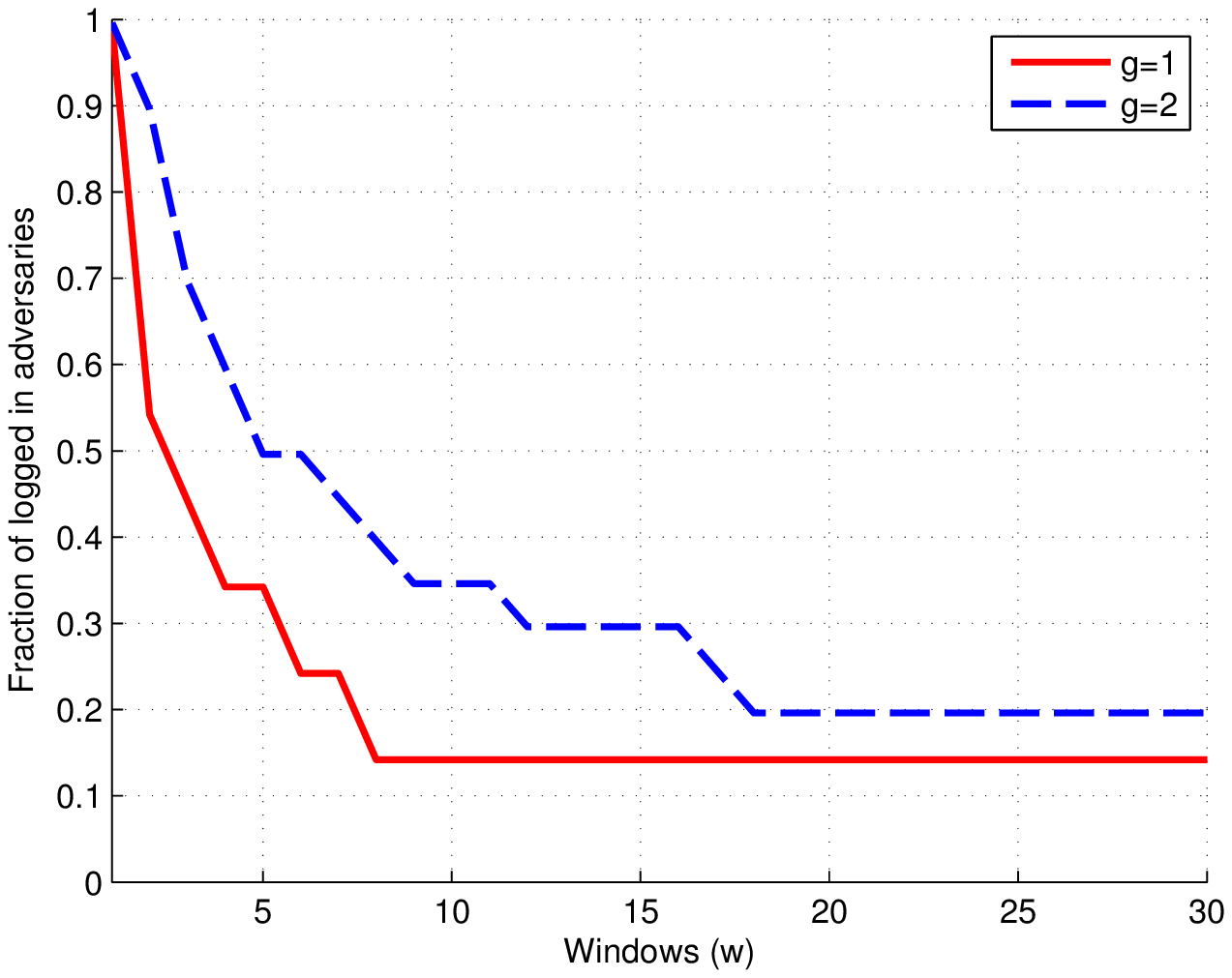}
		\caption{Fraction of attackers remaining logged in
                  after ($n$) authentication windows for different grace periods ($g$).}
		\label{fig:all_smart_time}
	\end{subfigure}
	\caption{\changeMika{Results for \textbf{opportunistic \allactivity} attackers. Opportunistic \allactivity attackers replicate easy mouse-hand movements of the victim.}}
\end{figure}
\fi
\ifllncs
(Appendix~\ref{app:extra_attackers})
\fi
summarize the performance of the \textbf{opportunistic \allactivity} attack. Compared to Figure~\ref{fig:kb_smart_fpr}, average FPR values in Figure~\ref{fig:all_smart_fpr} are somewhat worse for the attacker. This results in opportunistic \allactivity attackers being logged out at a higher rate compared to opportunistic \KBactivity 
attackers (although this is not statistically significant, with
$z=-1.082, r=-0.17$ and $p=0.279 > 0.05$). This is not surprising since mimicking all types of interactions is likely to be harder than mimicking typing interactions only. Nevertheless, opportunistic \allactivity attackers are somewhat more successful than na\"ive \allactivity attackers 
(but again not statistically significant, with $z=-1.514, r=0.24, p=0.130 > 0.05$). For example, with $g=1$, all na\"ive \allactivity attackers are logged out after 7 windows, while 25\% of the opportunistic \allactivity attackers succeed in remaining logged in.



\ifllncs

We also consider the question whether the inability of the attacker to see the victim hampers his ability to circumvent \zebra. Similar figures for FNR and fractions of logged in users in such an \textbf{audio-only opportunistic KB-only} attack are presented in Appendix~\ref{app:extra_attackers}. 
Such an attack is slightly less successful than an opportunistic \KBactivity attacker who is able to see his victim. 
It is also slightly less successful than a na\"ive \allactivity attack, but not significantly so.

Thus, we conclude that an attacker adopting an opportunistic approach can do better in circumventing \zebra than na\"ively mimicking all interactions. This may hold even when the attacker is hampered by not having visual access to the victim. An opportunistic \KBactivity attacker performs significantly better than a na\"ive \allactivity attacker.

\else

\begin{figure}[t!]
	\centering
	\begin{subfigure}[b]{\figwidth}
		\centering
		\includegraphics[width=0.7\figwidth]{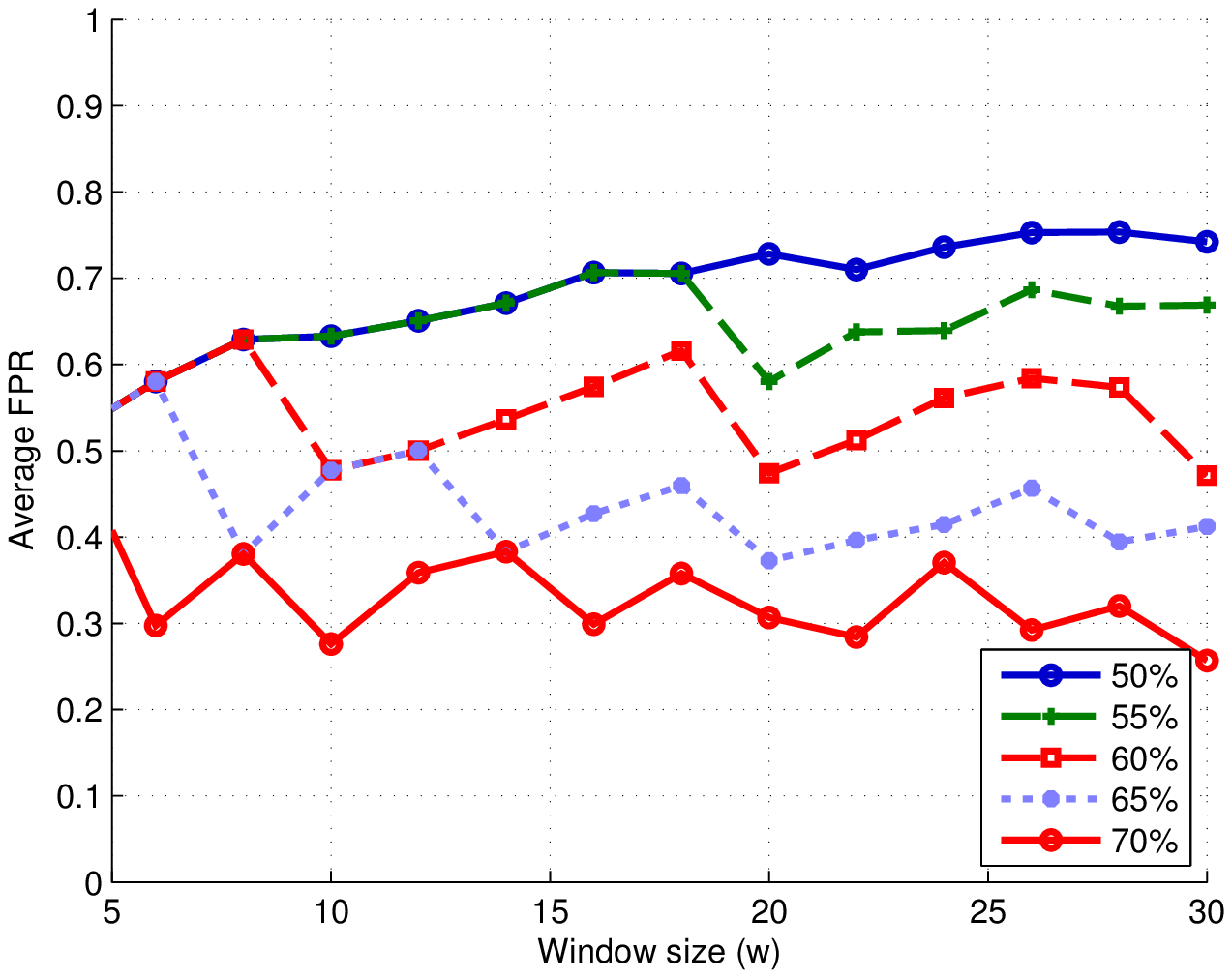}
		\caption{Average FPR for different threshold ($m$) values. \changeMika{Fraction of attacker windows that are classified as matching with the bracelet.}}
		\label{fig:kb_audioonly_fpr}
	\end{subfigure}

	\begin{subfigure}[b]{\figwidth}
		\centering
		\includegraphics[width=0.7\figwidth]{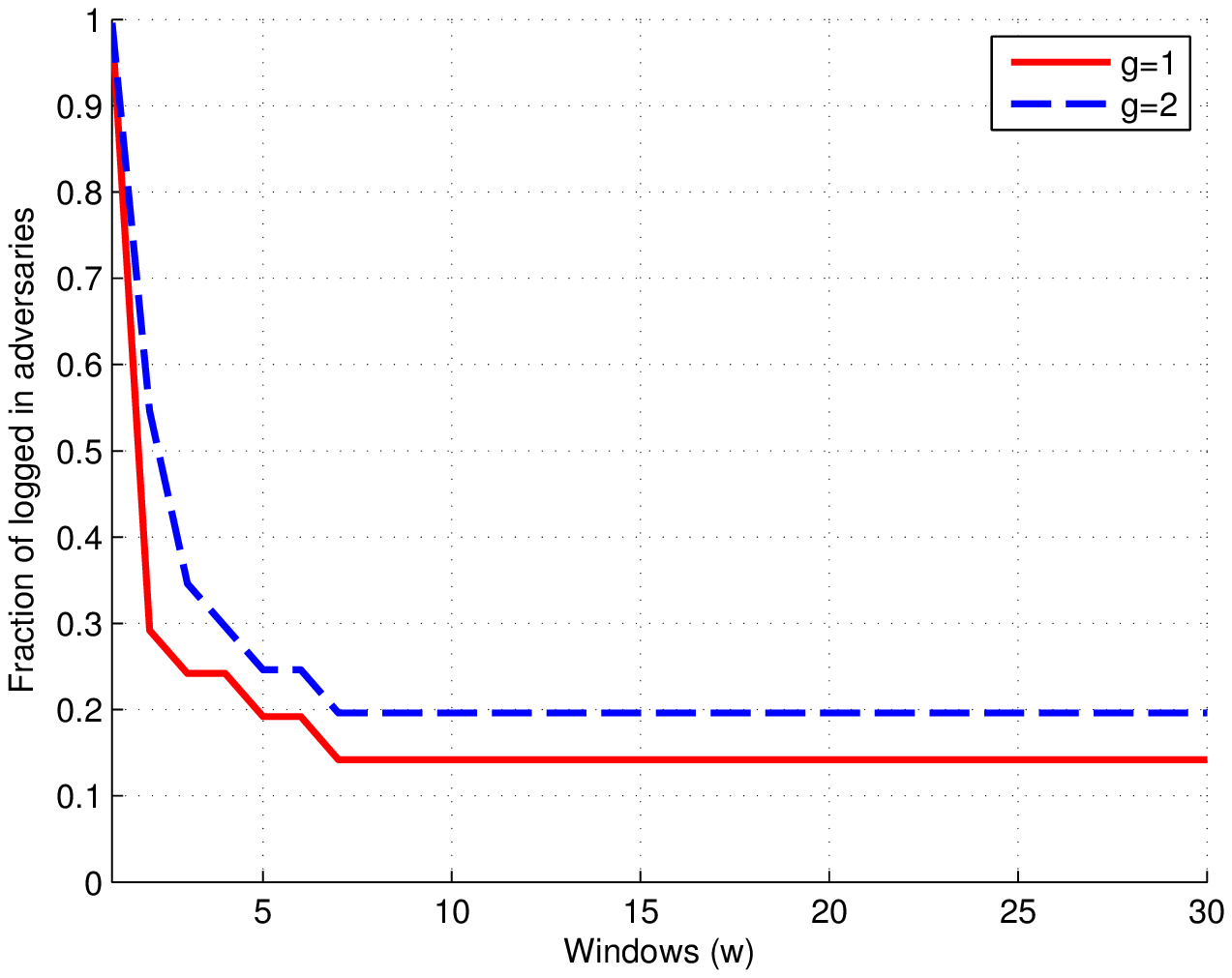}
		\caption{Fraction of attackers remaining logged in
                  after ($n$) authentication windows for different grace periods ($g$).}
		\label{fig:kb_audioonly_time}
	\end{subfigure}
	\caption{\changeMika{Results for \textbf{audio-only opportunistic \KBactivity} attackers. Audio-only opportunistic \KBactivity attackers eavesdrop on the victim and type only when they hear the victim typing.}}
\end{figure}

Next we consider the question whether the inability of the attacker to see the victim hampers his ability to circumvent \zebra. Figures \ref{fig:kb_audioonly_fpr} and \ref{fig:kb_audioonly_time} summarize the performance of an
\textbf{audio-only opportunistic \KBactivity} attacker. 
This attack is in line with prior attacks based on keyboard acoustic emanations \cite{Asonov04, Zhuang2009KAE}.
Prior attacks aimed at recognizing the keystrokes based on their sounds, while our attack attempts to recognize typing/mouse activities based on their sounds. One key difference is that our attack is manual, whereas prior attacks were automated (in fact, it seems that prior attacks can not be performed manually since a human attacker may not be able to distinguish between sounds of different keys). As such, our attack may be viewed as a new form of acoustic emanations attack targeted at the ZEBRA system.

Again, we see that such an attack is less successful than an opportunistic \KBactivity attacker who is able to see his victim. However, it is still more successful than a na\"ive \allactivity attack. Again, with $g=1$, 15\% of the audio-only opportunistic \KBactivity remain logged in after 6 windows.

Thus, we conclude that an attacker adopting an opportunistic approach can do better in circumventing \zebra than by na\"ively mimicking all interactions. This holds even when the attacker is hampered by not having visual access to the victim. An opportunistic \KBactivity attacker performs significantly better than a na\"ive \allactivity attacker.

\fi


%% file: chapters/discussion.tex


\section{Strengthening \zebra}
\label{subsec:strengthening}
Opportunistic attacks against \zebra succeed because of the
fundamental flaw in its design: it allows the adversary to control both interaction sequences Authenticator receives as input. First, the adversary has full control over the actual interaction sequence as he can choose the type and order of his terminal interactions. Second, he can indirectly influence the predicted interaction sequence as his terminal inputs cause Interaction Extractor to choose the times at which the victim's bracelet data is segmented and fed to the Interaction Classifier to generate the predicted interaction sequence. 

We can cast this as a general problem of \textit{tainted input}: accepting data which can be incorrect or outright malicious, and performing security-critical actions based on it. This is a common issue in any application or on-line service accepting input from potential adversaries. There are typically three counter-measures: (1) augmenting with trusted input, (2) marking untrusted input as tainted and performing taint tracking, or (3) sanitizing untrusted input before using it. As the sole purpose of the interaction sequences is authentication, taint tracking is not applicable in our case. Thus, we consider the other two potential solutions: using trusted input and input sanitization.

\vspace{1mm}
\noindent\textbf{Augmenting with Trusted Input}: Instead of allowing the terminal input to fully determine when Authenticator compares the two interaction sequences, a \textit{fundamental} fix is to base this determination additionally on bracelet data which is not under the control of the attacker. This would require inferring the predicted interaction sequence continuously from the bracelet data even when the terminal observes no actual interaction. If the predicted interaction sequence suggests that the user is interacting with a terminal, but no corresponding actual interaction is observed, Authenticator should output ``Different User''. This presupposes that the Interaction Classifier has very high precision \changeAsokan{(which we discuss below)}. 
Requesting data from the bracelet continually, rather than on demand, might lead to unwanted deauthentication if the event is not recognized.

Augmenting ZEBRA with Bluetooth proximity measurements means that we have another way of assurring ourselves that the user is nearby. We noticed that typical bluetooth signal strengths are within -5dB for users immediately close by, e.g. working at the terminal. Similarly, users walking nearby the terminal tend to have signals strengths within -15dB. Based on this, a three-level proximity calculation could be developed, classifying the proximity of the user as \textit{immediate}, \textit{near} or \textit{far} based on the Bluetooth signal strength. Users that are perceived as being near or far could have progressively increased authentication thresholds, e.g. increasing the threshold from 70\% to 80\% in case of near distance and further to 90\% in case of far distance. This would make mimicking attacks more difficult, because the attacker needs to be very close to the victim in order to have a lenient threshold.

\changeAsokan{In a centralized (multi-terminal) environment, it may be possible to use successful login events as an input for triggering deauthentication: a central system could recognize when a user logs into a terminal and automatically deauthenticate him) from any other terminal where he has an active logged-in session.}


\vspace{1mm}
\noindent\textbf{Sanitizing Untrusted Input}: Input sanitization can take the form of whitelisting (accepting specific well-formed inputs only) or blacklisting (rejecting a set of known malicious input patterns). Authenticator has two inputs that need to be sanitized: the actual interaction sequence and the predicted interaction sequence.

For example, one could attempt to prevent our opportunistic \KBactivity attacker by adopting a whitelisting approach of only accepting actual interaction sequences which contain {\em multiple types} of interactions, such as requiring periodic MKKM interactions interspersed with typing. However, since many legitimate user sessions can involve typing-only sequences, this remedy will violate the zero-effort requirement.

\zebra could also use blacklisting where certain types of input data can immediately trigger deauthentication. 
For example, if an input stream can reliably indicate the user standing up and walking away from the terminal, it can trigger deauthentication. Augmenting the bracelet data we currently use (accelerometer and gyroscope) with additional information, like heart-rate data available on many current smartwatches, can be used for this purpose. \changeMika{However, these fixes can seem privacy-invasive for some users.}

\vspace{1mm}
\noindent\textbf{Further Instances of Tainted Input}: We identify additional types of input interactions that an adversary can use to defeat \zebra. As the bracelet is assumed to be worn on the \changeMika{mouse-controlling} (e.g., right) hand, \zebra records an MKKM interaction after mouse activity only if it observes a keypress event on that side of the keyboard. This is done to reduce false negatives arising from a user who types with the \changeMika{keyboard-only} hand without removing his \changeMika{mouse-controlling} hand from the mouse. Again, such a design decision introduces a vulnerability: for example, in the case of a right handed victim, the attacker can type using only the left and middle parts of the keyboard (approximately 60\% of the keys) while the victim continues to use the mouse. Having previously recorded an interaction involving the mouse, \zebra will leave out all such subsequent typing from the segments it considers for comparison. This could be mitigated by blacklisting long typing sequences involving keys in the middle and non-dominant parts of the keyboard as such sequences are not typical in normal \changeMika{workstation} usage.

Another such vulnerability is when \attacker interacts by only moving and clicking the mouse. \changeMika{No event gets reported} \changeOtto{for these activities} \changeMika{and consequently an adversary can potentially do much harm, for example by copy-pasting words appearing in the screen.} Mare et al.~\cite{mare2014zebra} report that they did not consider mouse movement and click events as interactions because ``they did not contribute to \zebra's performance.'' However, including them would seem the most feasible defense. \changeMika{As we can see in our examples,} \changeAsokan{pre-mature optimization motivated by privacy (such as not collecting data under certain scenarios) may introduce security vulnerabilities.}

\vspace{1mm}
\noindent\textbf{\changeMika{Making the System Work in Real-time}}:
\changeMika{
It is well-known that on-line systems always bring new information of the usability, compared to off-line analysis. When we experimented with our end-to-end implementation in real-time, we noticed that} \changeAsokan{the original 3-class classifier (typing, scrolling and MKKM)} \changeMika{systematically} \changeAsokan{identified} \changeMika{bracelet} \changeAsokan{measurements} \changeMika{during walking and standing (``upright'')} \changeAsokan{as typing interactions. Similarly, measurements while the bracelet was simply} \changeMika{lying on the table (``idle'')} \changeAsokan{were classified as scrolling interactions}. \changeMika{The similarity between the} \changeAsokan{hand} \changeMika{movements in these} \changeAsokan{pairs of events} \changeMika{leads to similar magnitude-based features. Based on these observations, we} \changeAsokan{extended the original classifier to account for the new types of ``interactions'': idle and upright.} \changeAsokan{The new} \changeMika{classes were added as a post-processing step: a one hundred random tree ensemble was learned first (random forest), each tree contributing by voting between one of the three original classes. These} \changeAsokan{votes} \changeMika{were fed to a C4.5 decision tree, which learned decision thresholds for the five classes.
We noticed that} \changeAsokan{the performance of the resulting \textit{real-time system}} \changeMika{improved a great deal. 
We observed an overall improvement in} \changeAsokan{both} \changeMika{the accuracy and the ability} \changeAsokan{to generalize to} \changeMika{new users.} 

\vspace{1mm}
\changeMika{\noindent\textbf{Improving Machine Learning}:} \changeAsokan{Mare et al.~\cite{mare2014zebra} make use of accelerometers and gyroscopes that report measurements in three dimensions but use only magnitude values calculated from a single dimension. \zebra, and other similar techniques in general, can be extended to use measurements from all three dimensions.} \changeMika{The gravity component in individual axes can be eliminated with a low-pass filter \cite{sensorevent}. The added information from statistical measures in any individual direction can help in the discrimination of the classes, increasing the accuracy of the classifier in normal usage. With a better classifier we can raise the threshold ($m$) of authenticating a user interaction, lowering the FPR, while increasing the FNR. An acceptable FNR level can then be found as a compromise with receiver operating characteristic (ROC) curves, which shows the trade-off between TPR and FPR.}

\changeMika{The Scrolling events were more difficult to identify compared to others in our experiments (Table~\ref{fig:cm}). So improving the accuracy of these predictions is of interest.} \changeMika{Further feature engineering can increase the classification ability. Feature selection algorithms can select robust features that generalize the decision rules well. Feature selection can also help in increasing the battery life of the bracelet, since less information needs to be transmitted over Bluetooth from the bracelet to the computer for classification.}


\section{Discussion} 
\label{subsec:implementation_differences}
Despite our attempts to reproduce
the implementation described in \cite{mare2014zebra}, differences
remain. Although our implementation achieves lower FNR for legitimate
users, it incurs somewhat longer delays in logging out na\"ive
attackers. We were unable to reproduce the high rate of user interactions reported in \cite{mare2014zebra}. Despite these differences, the main result of our work holds because it is \textit{comparative}: we demonstrated that in our system, attackers adopting opportunistic strategies can significantly outperform a na\"ive \allactivity attacker. Such a comparative result will hold in any implementation of \zebra, including \cite{mare2014zebra}, despite any differences between implementations.

\vspace{1mm}
\changeMika{\noindent\textbf{Impact of Data Set}: }
\changeOtto{One contributor in performance differences could be a methodological difference we discovered with the original paper. The authors note that one user's ``wrist movement during keyboard and mouse interaction were very different compared to the other subjects", and one of the test users is logged out almost immediately. It is likely that a large fraction of this user's authentication windows are thus incorrectly classified, amounting to 1/20, i.e. 5 percent points difference in FPR between} \changeAsokan{their experiments and ours.}

\changeOtto{One potential explanation is that in \cite{mare2014zebra} only one of the users was left-handed, which may result in differences for this one user. The leave-one-user-out classifier training may also exaggerate this as it results in the classifier being trained with data from only right-handed users, but tested with data from the left-handed user. However, without access to the original test data, this cannot be verified.}

\vspace{1mm}
\noindent\textbf{Impact of Sampling Rate}: 
A notable difference lies in the sampling rate of the bracelet. We chose to use
commercial off-the-shelf smartwatches as bracelets because they are general-purpose devices readily available for a much larger audience and thus a realistic choice for deployment.
In such devices, the underlying sensor hardware limits the maximum sampling rate, typically 100-200 Hz on newer devices. Our LG smartwatch supported a sampling rate of 200 Hz. This is less than the 500Hz special-purpose Shimmer bracelet used in~\cite{mare2014zebra}.

The choice of sampling rate has an impact on power consumption~\cite{DBLP:conf/huc/BrajdicH13}. 
On Android, the sampling rate can be set to lower levels to save energy at the cost of reduced accuracy. 
\changeMika{The features we collect are mostly statistical measures calculated from the distribution of magnitude values measured during the event and should be quite stable as long as there are enough data points to calculate the values from. }

\changeMika{To evaluate the effect of sampling rates, we collected a small dataset from normal computer usage with 200 Hz sampling rate. We downsampled it to 100 Hz, 50 Hz and 25 Hz data sets by passing every second, fourth or eight measurement signal to Segmenter. 
The datasets were generated using the same data: the number of features and the number of events are the same at all frequencies, but the number of measurement signals used to calculate the features were different. 
We noticed that some features (e.g. skewness) could frequently not be calculated for short events at low frequencies because Segmenter could not pass enough measurements signal values to Feature Extractor. Sample skewness needs at least three values to be calculated. As a rule of thumb, the frequency of the bracelet needs to be at least $f_{min} = s_{min} / d_{min}$ to catch enough measurements for feature calculation, when $s_{min}$ (3) is the minimum amount of signal measurements needed to calculate all features and $d_{min}$ (25 ms) is the minimum duration of a classifiable event. With our end-to-end system parameter settings this would be $f_{min} = 120$ Hz.}
\changeMika{For devices operating at lower sampling rates, the minimum acceptable duration of interactions should be increased accordingly.}

\changeMika{Typically lower sampling rates increase the noise in features, which in turn changes class boundaries. We expect minor classes to get misclassified as major classes more frequently. In the worst scenario, everything gets classified as the major class (typically this would be typing in our scenarios). Lower sampling rates would increase the FPR in this way. This is not the case in 200 Hz, as can be seen in Table~\ref{fig:cm} (Section~\ref{sec:performance_evaluation}).}
We experimented \changeMika{on our real-time system} with one such lower rate (20 Hz), at which the traces contain very little information for each interaction causing \zebra to become insensitive to synchronization delays as long as 1s. 
At higher rates above \changeMika{120 Hz} there was no noticeable difference. Therefore, we conclude that while \zebra is unreliable at very low sampling rates, its performance \changeMika{was found to be} steady at or above \changeMika{120 Hz}. 




%% file: chapters/related_work.tex
\section{Related Work}

User authentication is commonly based on three different factors: something you know, something you have, and something you are. Many traditional authentication methods rely on the first two. Passwords still remain very popular, and they are often complemented with some sort of physical tokens, such as RSA SecurID~\cite{securid}. The downside is the need to deploy and carry these tokens. 

A large body of research has considered biometric authentication. Examples include the use of fingerprint~\cite{clancy2003secure}, hand~\cite{kumar2009personal}, iris~\cite{chong2005iris}, facial~\cite{beumier2000automatic} or blood vessel information~\cite{watanabe2005palm}. Biometric authentication is attractive because they reduce the user burden by removing the need to memorize secrets or having to carry external tokens. However, these schemes can still be vulnerable to spoofing, and introduce new issues such as the problem of revocation and raise privacy concerns. Also, traditional biometric authentication is not transparent to the user.

The desire to minimize the user burden of authentication has led to a quest for transparent and continuous authentication schemes that can be ``zero-effort.'' One approach uses proximity-based authentication where the presence of a personal device is used to authenticate the user. Such schemes can be based on RFID, NFC, Bluetooth or even WiFi signal strength. The appeal is the possibility to use existing devices seamlessly, but unfortunately the drawbacks include limited accuracy and vulnerability to spoofing and replay attacks~\cite{hancke2005practical}~\cite{hancke2006practical}~\cite{francis2010practical}.

Behavioral biometrics consider behavior intrinsic to specific individuals. A common example is gait, identified from video or acceleration information. Gafurov et al.~\cite{gafurov2006biometric} perform authentication based on a user's gait, which is characterized by recorded accelerations from a hip-worn device. The same author also considers~\cite{gafurov2007spoof} spoofing attacks against gait-based authentication.
A subset of these behavioral biometrics are keystroke and typing based authentication schemes. In an early work, Joyce et al.~\cite{joyce1990identity} and Monrose et al.~\cite{monrose2000keystroke}~\cite{monrose2002password} identify users based on their typing rhythm. They consider the inter-key latencies and are able to effectively authenticate users. More recently, Ahmed et al. consider mouse dynamics for authentication~\cite{ahmed2007new}. ~\cite{de2012touch} distinguishes users based on how they input touch patterns into a smartphone. However, Tey et al.~\cite{tey2013can} show how through training, attackers can learn to defeat keystroke biometrics based authentication.

Combining multiple types of transparent authentication schemes, such as the proposal by Riva et al.~\cite{DBLP:conf/uss/RivaQSL12}, can improve the overall performance. But the design of such systems is complex and remains an open research problem.\\

%% file: chapters/conclusions.tex
\section{Conclusions}

\changeMika{\zebra is an interesting and useful approach as a 
zero-effort deauthentication system. We identified a subtle design flaw
in this approach, which is (1) easier for the human operators to 
perform and (2) more robust, compared to the na\"ive attacks
studied by the authors of the \zebra scheme \cite{mare2014zebra}.}
We demonstrated that a malicious adversary who adopts an
opportunistic strategy can defeat \zebra. This is at odds with the
positive results reported in \cite{mare2014zebra} but is explained by
their attackers using a na\"ive strategy of trying to mimic all
interactions of a victim. Our attack is done in a
typical usage scenario. While physical mitigations, such as visual
barriers, might make our specific attack less successful, the
underlying vulnerability still stands.  Although susceptible to
opportunistic adversaries, \zebra still performs well against
accidental misuse by innocent adversaries, which is possibly the most likely threat in scenarios that \zebra was originally designed for. However, systems are often used in contexts that the designers did not originally envisage. Therefore, we believe that recognizing the limits of the original design of \zebra against malicious adversaries is the first step towards strengthening its resistance so that it can be used in scenarios where malicious adversaries pose a significant threat.
The approaches we
identified in Section~\ref{subsec:strengthening} can help secure
\zebra without losing its desirable properties. We are developing
these approaches further in our current work.
\changeAsokan{More generally, we showed that subtle design assumptions based on premature usability and privacy considerations can adversely impact security of a system. We also highlight the importance of ensuring that adversary models used in analyzing the security of systems are realistic and do not underestimate attacker capabilities.}


%% file: chapters/appendices.tex
\newpage
\appendix

\section{Parameters}
\label{app:parameteres}

The parameters we use in our \changeMika{end-to-end} system are listed in Table \ref{tab:parameters}.
\begin{table}[!h]
\centering
	\caption{Parameters and their values used in this paper.}
\begin{minipage}{\columnwidth}
	\begin{tabularx}{\textwidth}{|X|X|X|}
	\hline
	\textbf{Parameter} & \textbf{Value} & \textbf{Rationale}\\[6pt]
	\hline
	Min. duration\footnote{For scrolling, also a minimum of 5 recorded events.} & 25 ms & \cite{mare}\\
	\hline
	Max. duration\footnote{\label{fn:two}For MKKM, a max. duration and idle threshold of 5s.\cite{mare} } & 1 s & \cite{mare2014zebra}\\
	\hline
	Idle threshold\textsuperscript{\ref{fn:two}} & 1 s & \cite{mare}\\	
	\hline
	Window size ($w$) & 5-30 & \cite{mare2014zebra}\\
	\hline
	Match threshold ($m$) & 50-70\% & \cite{mare2014zebra}\\
	\hline
	Overlap fraction ($f$) & 0 & Estimated\footnote{Estimate based on reported \cite{mare2014zebra} times \& authentication windows needed for logging out users.}\\
	\hline
	Grace period ($g$) & 1-2 & \cite{mare2014zebra}\\
	\hline
	\end{tabularx}
\end{minipage}
	\label{tab:parameters}
\end{table}

\section{Features}
\label{app:features}
We consider the same features as in \cite{mare2014zebra}, listed in Table \ref{tab:features}. These are extracted from segments of sensor readings and used to classify interactions.
\begin{table}[t!]
\centering
	\caption{Features used in this paper.}
	\begin{tabular}{|l|l|}
	\hline
	\textbf{Feature} & \textbf{Description} \\[6pt]
	\hline
	Mean & mean value of signal\\
	\hline
	Median & median value of signal\\
	\hline
	Variance & variance of signal\\	
	\hline
	Standard Deviation & standard deviation of signal\\
	\hline
	MAD & median absolute \changeMika{deviation}\\
	\hline
	IQR & inter-quartile range\\
	\hline
	Power & power of signal\\
	\hline
	Energy & energy of signal\\
	\hline
	Peak-to-peak & peak-to-peak amplitude\\
	\hline
	Autocorrelation & similarity of signal\\
	\hline
	Kurtosis & peakedness of signal\\
	\hline
	Skewness & asymmetry of signal\\
	\hline
	\end{tabular}
	\label{tab:features}
\end{table}

\ifllncs
\section{Deauthentication Time}
\label{app:deauthentication_time}

Figure~\ref{fig:kb_smart_time_sec} represents the fraction of logged
in \KBactivity attackers as a function of time (in minutes). 
Figure~\ref{fig:kb_smart_time_sec} corresponds to Figure~\ref{fig:kb_smart_time} in Section~\ref{sec:results}.

\begin{figure}[h]
  \centering
  \includegraphics[width=\textwidth]{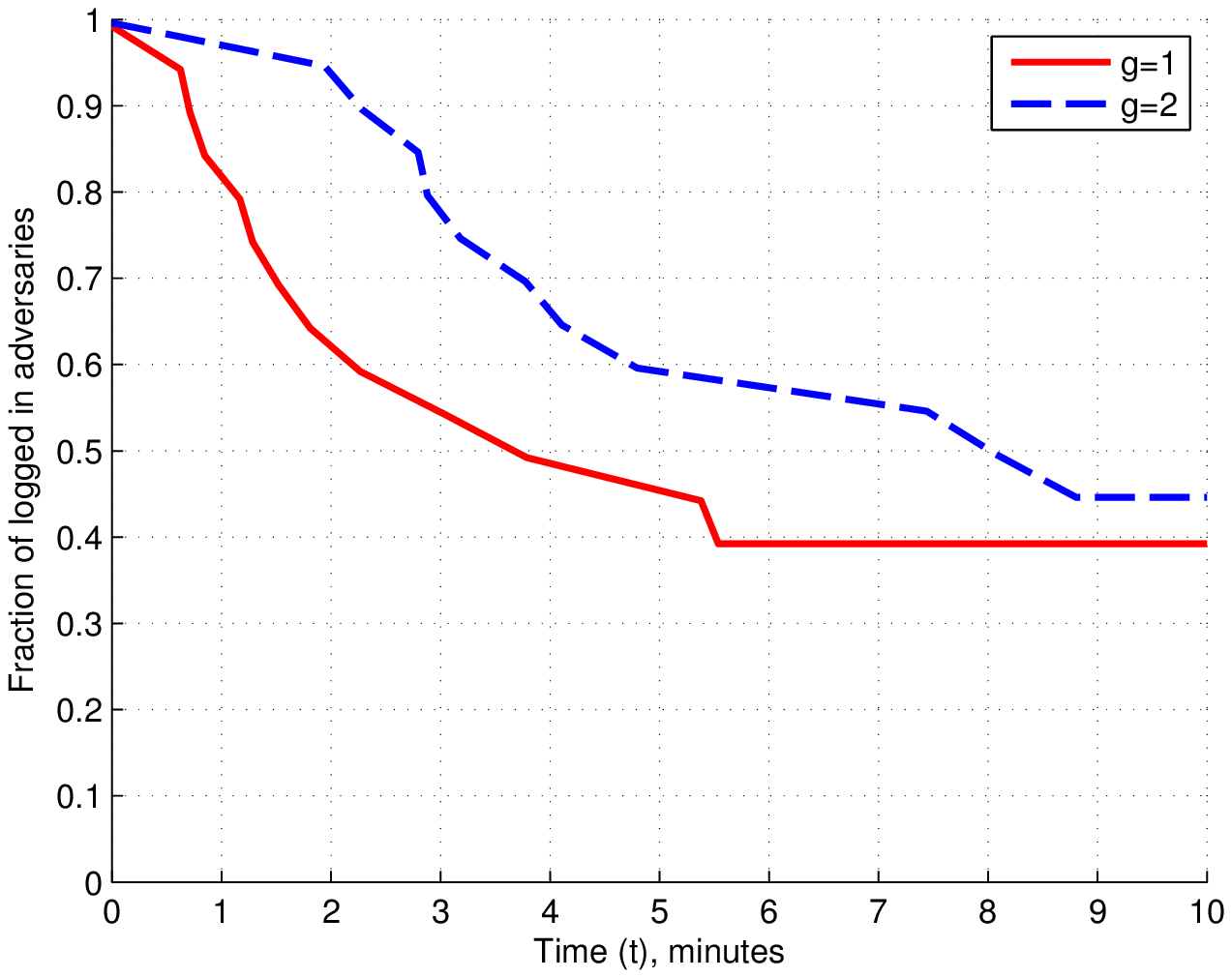}
\caption{\textbf{Opportunistic \KBactivity} attacker: 
  Fraction of attackers remaining logged in after ($t$) minutes
  (with $w=20, m=60\%)$, for different grace periods ($g$).}
  \label{fig:kb_smart_time_sec}
\end{figure}

\vspace{-5mm}
\section{Opportunistic all-activity and audio-only attackers}
\label{app:extra_attackers}
In opportunistic \allactivity attack, \attacker's selection
criterion is the likelihood of correctly mimicking \victim. In
particular, \attacker will use the following guidelines:
\begin{itemize}
\itemsep0em
\item Once \attacker successfully mimics a keyboard to mouse interaction, he
  is free to carry out any interaction involving the mouse (scroll,
  drag, move) at will because the bracelet measurements for all
  activities involving the mouse are likely to be similar.
\item If \attacker fails to quickly mimic a keyboard to mouse (or vice
  versa) interaction, he
  does nothing until the next opportunity for an MKKM
  interaction arises (foregoing all interactions until after the MKKM
  is completed).
\end{itemize}

\begin{figure}[htbp]
\centering
\begin{subfigure}{.5\textwidth}
  \centering
  \includegraphics[width=1\textwidth]{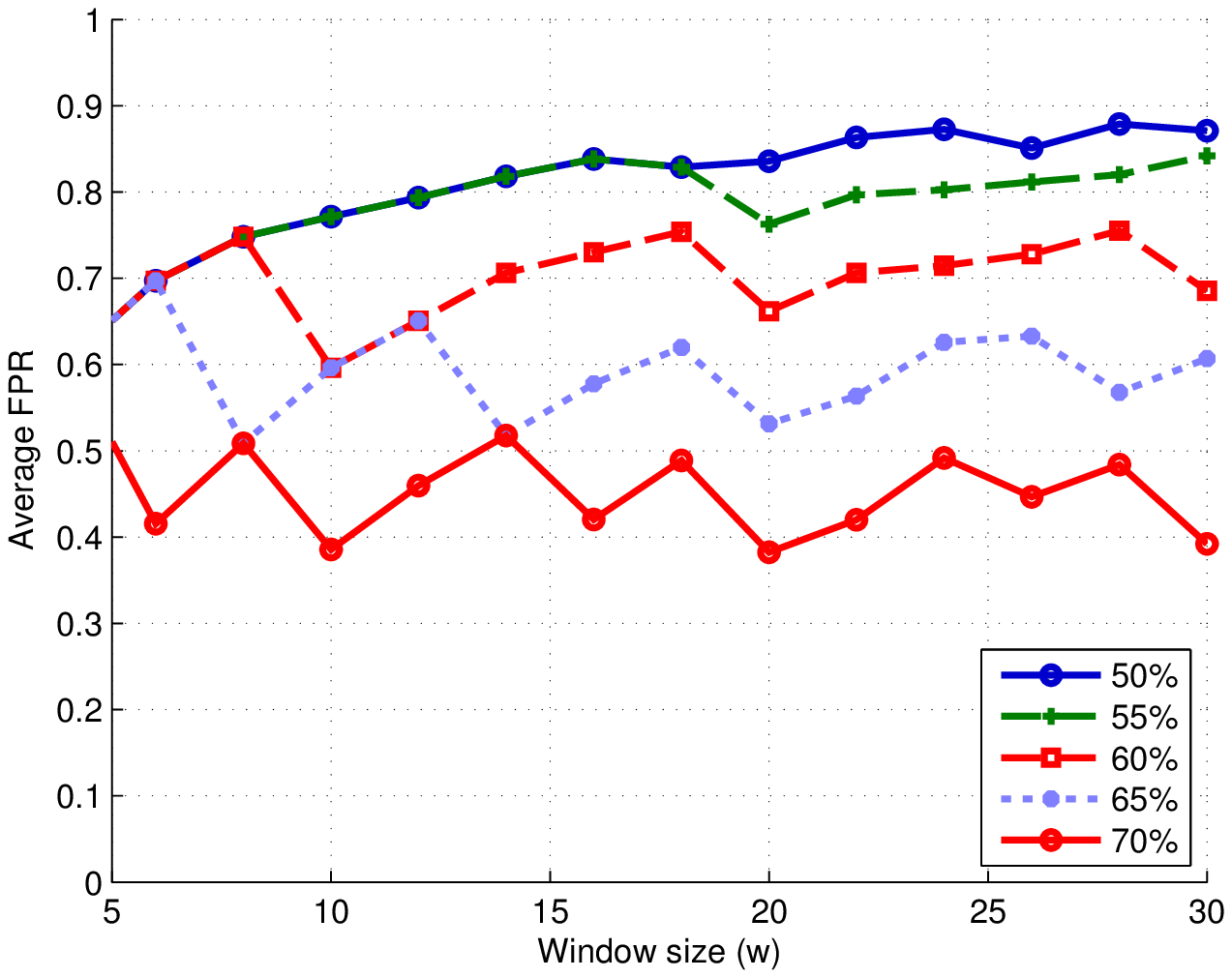}
  \caption{}
  \label{fig:all_smart_fpr}
\end{subfigure}%
\begin{subfigure}{.5\textwidth}
  \centering
  \includegraphics[width=1\textwidth]{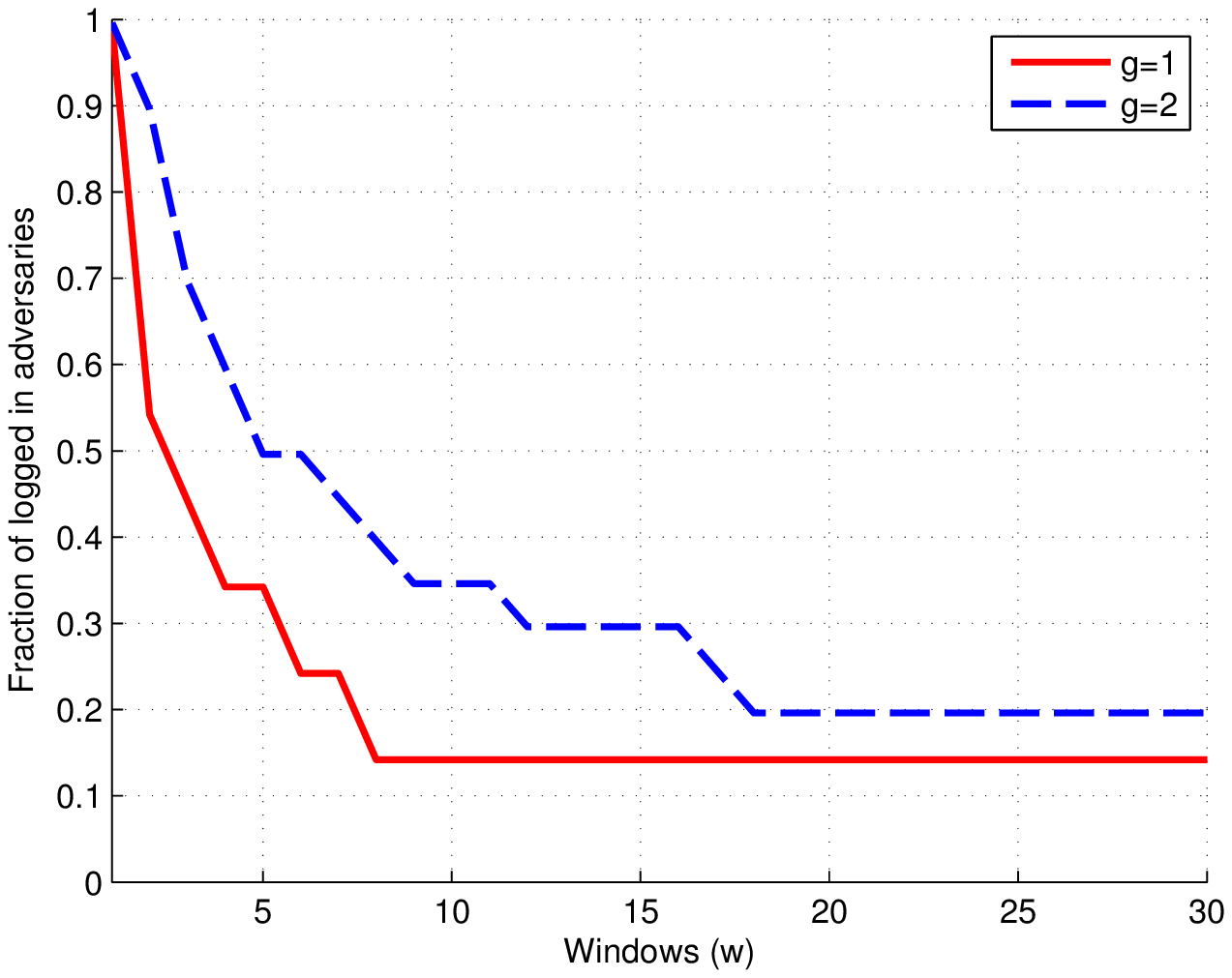}
  \caption{}
  \label{fig:all_smart_time}
\end{subfigure}
\caption{\textbf{Opportunistic \allactivity} attacker: \textbf{(a)}
  Average FPR for different threshold ($m$) values. \textbf{(b)}
  Fraction of attackers remaining logged in after ($n$) authentication
  windows (with $w=20, m=60\%)$, for different grace periods ($g$).}
\centering
\begin{subfigure}{.5\textwidth}
  \centering
  \includegraphics[width=1\textwidth]{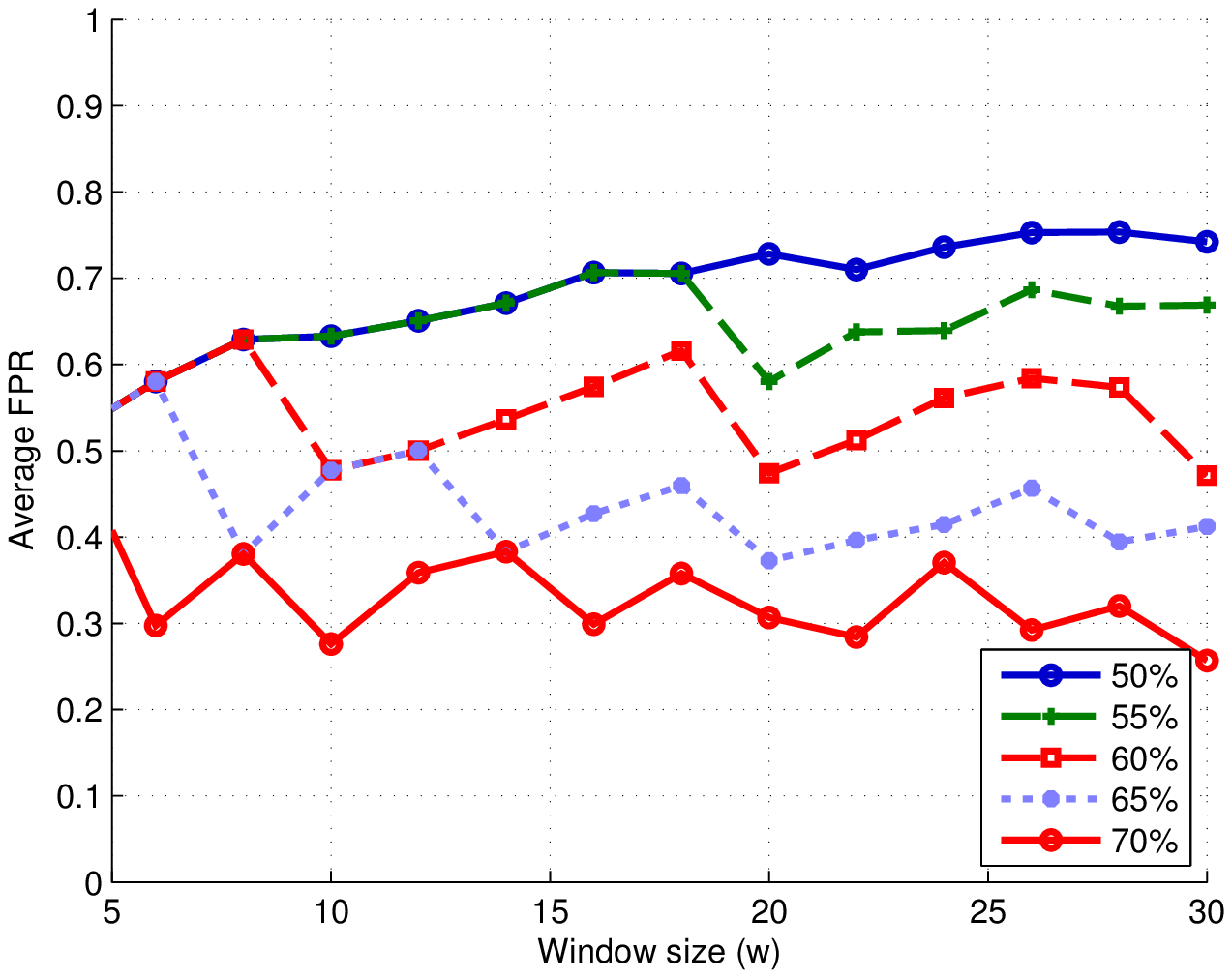}
  \caption{}
  \label{fig:kb_audioonly_fpr}
\end{subfigure}%
\begin{subfigure}{.5\textwidth}
  \centering
  \includegraphics[width=1\textwidth]{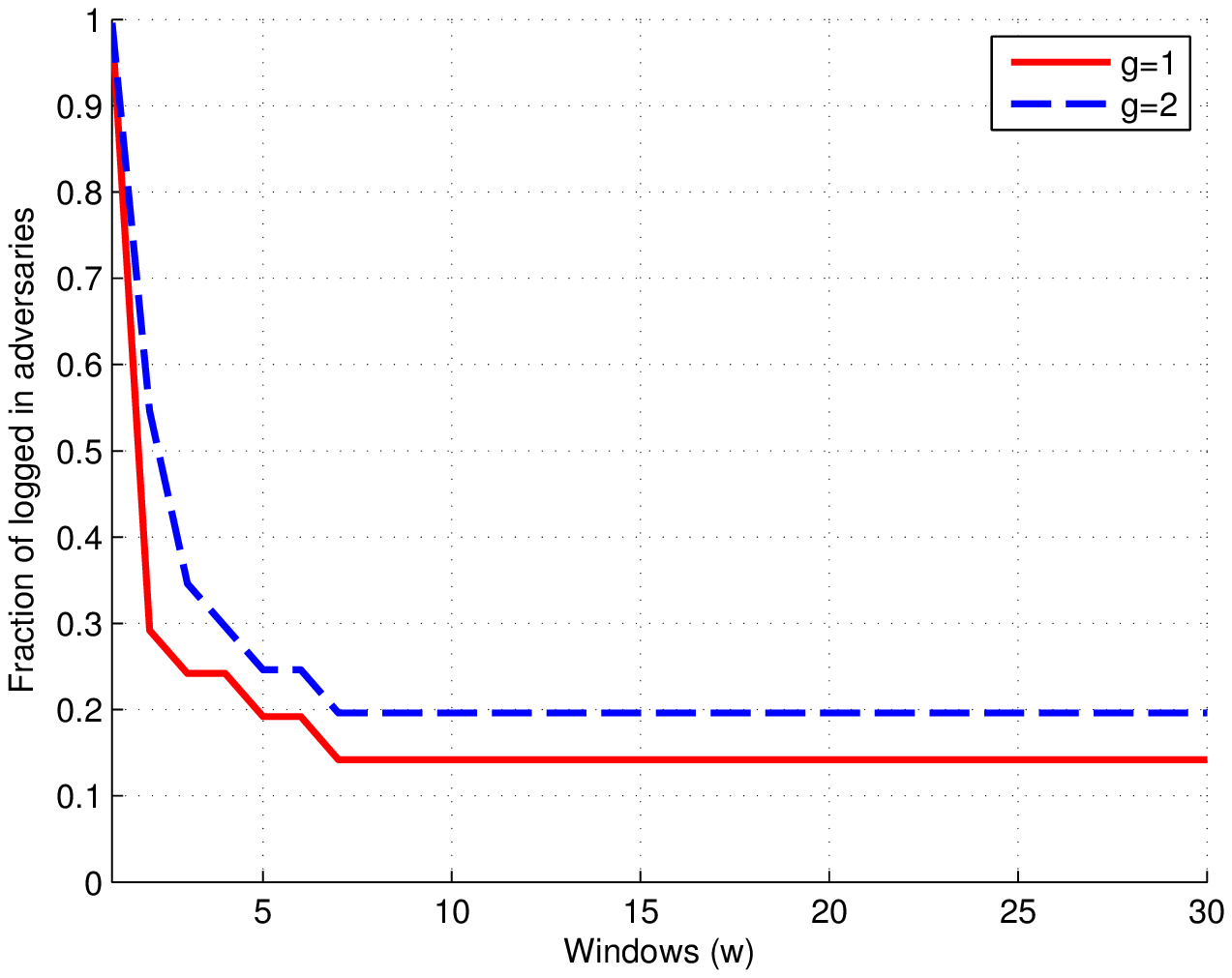}
  \caption{}
  \label{fig:kb_audioonly_time}
\end{subfigure}
\caption{\textbf{Audio-only opportunistic \KBactivity} attacker:
  \textbf{(a)} Average FPR ($w$) for different threshold ($m$)
  values. \textbf{(b)} Fraction of attackers remaining logged in after
  ($n$) authentication windows (with $w=20, m=60\%)$, for different grace periods ($g$).}
\end{figure}
\fi